\documentclass[11pt,titlepage,a4paper]{article}
\usepackage{bozhomac}	% loads: full AMSLaTeX and AMSFonts
\usepackage{bozhlogo}	% the logo of BOZHO command \BOZHO
\usepackage{cite}	% for condensing sequential cited works
\usepackage{tabularx}	% tables with automatically calculated column widths
\usepackage{varioref}	% flexible cross-references

\title{\bfseries	\vspace*{-2.13in}\enlargethispage{2ex}
{\huge Normal frames and\\ linear transports along paths\\ in vector bundles}
}

\vspace{1.1ex}

\author{
Bozhidar Z. Iliev
\thanks{Laboratory of Mathematical Modeling in Physics,
Institute for Nuclear Research and \mbox{Nuclear} Energy,
Bulgarian Academy of Sciences,
Boul. Tzarigradsko chauss\'ee~72, 1784 Sofia, Bulgaria}
\thanks{E-mail address: bozho@inrne.bas.bg}
\thanks{URL: http://theo.inrne.bas.bg/$\sim$bozho/}
}

\date{
 \vspace{1.5ex}%
	\ShortTitle{Normal frames and linear transports along paths}\\[0.27ex]
 \vspace{3.2ex}
	\begin{tabular}{r@{$\colon\to~$}l}
 \vspace{0.09ex} Basic ideas	& April 4 (10.16 a.m.), 1998	\\[0.09ex]
 \vspace{0.09ex} Began		& April 20, 1998 		\\[0.09ex]
 \vspace{0.09ex} Ended		& May 24 (9.07 a.m.), 1998 	\\[0.09ex]
 \vspace{0.09ex} Initial Typeset& June 4--19, 1998	\\[0.09ex]
 \vspace{0.09ex} Revised	& Jan \& Jun \& Oct, 1999; Jan--Feb, 2000\\[0.09ex]
 \vspace{0.09ex} Revised	& Feb \& Dec, 2002; January 2005
								\\[0.09ex]
 \vspace{0.09ex} Last updated	& March 7, 2005 	\\[0.09ex]
 \vspace{0.27ex} Produced	& \fbox{\today}	\\[0.27ex]
	\end{tabular} \\[1.27ex]
	\begin{tabular}{r@{$\colon~$}l}
 \vspace{0.27ex} http://www.arXiv.org e-Print archive No. & gr-qc/9809084
								\\[0.27ex]
	\end{tabular} \\[-0.27ex]
 \vspace{2.ex}{\Huge\BOZHO}	\\[2.ex]
\vspace{0.27ex}\Subject{Differential geometry} \\[2.27ex]
	\begin{tabular}{r@{\hspace{0.512em}}|@{\hspace{0.512em}}l}
\vspace{0.27ex}
&
\vspace{0.27ex}%\\[0.27ex]
	\end{tabular} \\[1.27ex]
\vspace{0.27ex}\\[0.27ex] }

\begin{document}		% BEGINNING OF THE DOCUMENT

\renewcommand{\thepage}{\roman{page}}

\renewcommand{\thefootnote}{\fnsymbol{footnote}} % special footnote symbols
\maketitle				% the title (page) is put here
\renewcommand{\thefootnote}{\arabic{footnote}}   % usual footnote symbols

\tableofcontents		% the table of contents is put here

%%%%%%%%%%%%%%%%%%%%%%%%%%%%%%%%%%%%%%%%%%%%%%%%%%%%%%%%%%%%%%%%%%%%%%%%%%%%%
%%%%%									%%%%%
%%%%%		actual beginning of the document			%%%%%
%%%%%									%%%%%
%%%%%%%

%%%%%%%%%%%%%%%%%%%%%%%%%%%%%%%%%%%%%%%%%%%%%%%%%%%%%%%%%%%%%%%%%%%%%%

\begin{abstract}
	The theory of linear transports along paths in vector bundles,
generalizing the parallel transports generated by linear connections, is
developed. The normal frames for them are defined as ones in which their
matrices are the identity matrix or their coefficients vanish. A number of
results, including theorems of existence and uniqueness, concerning normal
frames are derived. Special attention is paid to the important case when the
bundle's base is a manifold.  The normal frames are defined and investigated
also for derivations along paths and along tangent vector fields in the last
case. It is proved that normal frames always exist at a single point or along
a given (smooth) path. On other subsets normal frames exist only as an
exception if (and only if) certain additional conditions, derived here, are
satisfied. Gravity physics and gauge theories are pointed out as
possible fields for application of the results obtained.

\end{abstract}

%%%%%%%%%%%%%%%%%%%%%%%%%%%%%%%%%%%%%%%%%%%%%%%%%%%%%%%%%%%%%%%%%%%%%%%%%
\renewcommand{\thepage}{\arabic{page}}
%%%%%%%%%%%%%%%%%%%%%%%%%%%%%%%%%%%%%%%%%%%%%%%%%%%%%%%%%%%%%%%%%%%%%%%%%

\section {Introduction}
\label{4-Introduction}

	Conventionally, local coordinates or frames (or frame fields), which
can  be holonomic or not, are called \emph{normal} if in them the coefficients
of a linear connection vanish on some subset, usually a submanifold, of a
differentiable manifold. Until recently the existence of normal frames was
known (proved) only for symmetric linear connections on submanifolds of a
(pseudo\ndash ) Riemannian
manifold~\cite{Schouten/physics,Fermi,Veblen,Schouten/Ricci,ORai}.  New
light on these problems was thrown in the series of
papers~\cite{bp-Frames-n+point,bp-Frames-path,bp-Frames-general} where a
comprehensive analysis of the normal frames for derivations of the tensor
algebra over a differentiable manifold is given; in particular they completely
cover the exploration of normal frames for arbitrary linear connections on a
manifold. These strict results are applied in~\cite{bp-PE-P?} for rigorous
analysis of the equivalence principle. This results in two main
conclusions: the (strong) equivalence principle (in its `conventional'
formulations) is a provable theorem and the normal frames are the
mathematical realization of the physical concept of `inertial' frames.
Another physical application the normal frames find is in the bundle
formulation of quantum mechanics~\cite{bp-BQM-introduction+transport}. In
this approach the normal frames realize the (shift to the) bundle Heisenberg
picture of motion~\cite{bp-BQM-pictures+integrals}.

	The present investigation is a completely revised and expanded
version of~\cite{bp-LTP-general}. It can also be considered as a continuation
of the series of works~\cite{bp-Frames-n+point,bp-Frames-path,bp-Frames-general}
which are its special cases and, at the same time, its supplement. Here we
study a wide range of problems concerning frames normal for linear transports
and derivations along paths in vector bundles and for derivations along
tangent vector fields in the case when the bundle's base is a differentiable
manifold. In the last case, the only general result known to the author and
concerning normal frames is~\cite[p.~102, theorem~2.106]{Poor}.

	The structure of this work is as follows.

	Sect.~\ref{4-Sect2} is devoted to the general theory of linear
transports along paths in vector fibre bundles which is a far-reaching
generalization of the theory of parallel transports generated by linear
connections.%
\footnote{%
This result is not explicitly proved here. The interested reader is referred
to~\cite{bp-TP-parallelT} for details and the proof of this assertion.%
}
The general form and other properties of these transports are studied. A
bijective correspondence between them and derivations along paths is
established. In Sect.~\ref{4-Sect3}, the normal frames are defined as ones in
which the matrix of a linear transport along paths is the unit (identity)
one or, equivalently, in which its coefficients, as defined in
Sect.~\ref{4-Sect2}, vanish `locally'. A number of properties of normal
frames are found. In Sect.~\ref{4-Sect4} is explored the problem of existence
of normal frames. Several necessary and sufficient conditions for such
existence are proved and the explicit construction of normal frames, if any,
is presented.

	Sect.~\ref{4-Sect5} concentrates on, possibly, the most important
special case of frames normal for linear transports or derivations along
smooth paths in vector bundles with a differentiable manifold as a base.
A specific necessary and sufficient condition for existence of normal frames
in this case is proved in Sect.\ref{4-Subsect5.1}. In particular, normal frames
may exist only for those linear transports or derivations along paths whose
(2\Ndash index) coefficients linearly depend on the vector tangent to the path
along which they act. Obviously, this is a generalization of the derivative
along curves assigned to a linear connection. Sect.~\ref{4-Sect6} is devoted
to problems concerning frames normal for derivations along tangent vector
fields in a bundle with a manifold as a base. Necessary and sufficient
conditions for the existence of these frames are derived. The conclusion is
made that there is a one\nobreak-to\nobreak-one onto correspondence between the
sets of linear transports along paths, derivations along paths, and
derivations along tangent vector fields all of which admit normal frames.

	Sect.~\ref{4-Sect12} concerns a special type of normal frames in
which the 3\ndash index coefficients, if any, of a linear transport along
paths vanish.

	In Sect.~\ref{4-Conclusion} are presented some general remarks.
It is shown that the results
of~\cite{bp-Frames-n+point,bp-Frames-path,bp-Frames-general} remain valid,
practically without changes, for (strong) normal frames in vector bundles
with a manifold as a base.

% 	In the~\ref{4-SectE}, the developed formalism is
% specialized to the simple case of a one\ndash dimensional vector bundle over
% a manifold and the results obtained are applied to the bundle description of the
% classical electromagnetic field. In particular, the inertial frames for it
% are discussed.

\vspace{1.4ex}

	All fibre bundles in this work are vectorial ones. The
base and total bundle space of such bundles can be general topological
spaces. However, if some kind of differentiation in one/both of these spaces
is required, it/they should possess a smooth
structure; if this is the case, we require it/they to be smooth, of class
$C^1$, differentiable manifold(s). Starting from Sect.~\ref{4-Sect5}, the base
and total bundle space are supposed to be $C^1$ manifolds.
Sections~\ref{4-Sect2}--\ref{4-Sect4} do not depend on the existence of a
smoothness structure in the bundle's base. Smoothness of the bundle space is
partially required in sections~\ref{4-Sect2}--\ref{4-Sect4}.%
\footnote{~%
The bundle space is required to be a $C^1$ manifold in
Sect~\ref{4-Sect2} (starting from
definition~\ref{4-Defn2.2}), in definition~\ref{4-Defn3.1'},
in propositions~\ref{4-Prop3.1}--\ref{4-Cor3.1*},
if~\eref{4-3.1c} and~\eref{4-3.1d} are taken into account, in
theorem~\ref{4-Thm3.2}, and in proposition~\ref{4-Prop4.4}.%
}

\section {Linear transports along paths in vector bundles}
\label{4-Sect2}
\index{linear transport along paths|(}

	 From different view\ndash points, the connection theory can be found
in many works, like%
~\cite{Kobayashi-1957,K&N-1,Warner,Steenrod,Husemoller,
Greub&et_al.-1,Mangiarotti&Sardanashvily}.
As pointed in these and many other references, the concept of a parallel
transport is defined on the base of the one of a connection. The opposite
approach, \ie the definition of a connection on the ground of an axiomatically
defined  concept of a parallel transport, is also know and considered
in~\cite{Lumiste-1964,Lumiste-1966,Teleman,Dombrowski,Lumiste-1971,
Mathenedia-4,Durhuus&Leinaas, Khudaverdian&Schwarz,Ostianu&et_al.,Nikolov,
Poor}.

	The purpose of the present section is an introduction and partial study
of an axiomatic definition (and generalization) of parallel transport in
vector bundles, called \emph{transport along paths} which in the particular
case is required to be linear.

	Our basic definition of a (linear) transport along paths is
definition~\ref{4-Defn2.1} below. Comparing it
with~\cite[definition~2.1]{bp-LT-Deriv-tensors} and taking into
account~\cite[proposition~4.1]{bp-LT-Deriv-tensors}, we conclude that special
types of general linear transports along paths are: the parallel
transport assigned to a linear connection (covariant derivative) of the
tensor algebra of a
manifold~\cite{K&N-1,Schouten/Ricci},
Fermi\ndash Walker transport~\cite{Hawking&Ellis,
Synge}, Fermi transport~\cite{Synge},
Truesdell transport~\cite{Walwadkar,Walwadkar&Virkar},
Jaumann transport~\cite{Radhakrishna&et_al.},
Lie transport~\cite{Hawking&Ellis,Schouten/Ricci},
the modified Fermi\ndash Walker and Frenet\ndash Serret
transports~\cite{Dandoloff&Zakrzewski},
etc.
Consequently, definition~\ref{4-Defn2.1} is general enough to cover a list of
important transports used in theoretical physics and mathematics. Thus
studying  the properties of the linear transports along paths, we can make
corresponding conclusions for any one of the transports mentioned.%
\footnote{%
The concept of linear transport along paths in vector bundles can be
generalized to the transports along paths in arbitrary
bundles~\cite{bp-TP-general} and to transports along maps%
\index{transport along maps|nn}
in bundles~\cite{bp-TM-general}. An interesting consideration of the concept
of (parallel) `transport' (along closed paths) in connection with homotopy
theory and the classification problem of bundles can be found
in~\cite{Stasheff-PT}. These generalizations are out of the scope of the
present work.%
}

	As we said above, definition~\ref{4-Defn2.1} below realizes, an
axiomatic approach to the concept of a parallel
transport~\cite{Lumiste-1964,Teleman,Dombrowski,Lumiste-1971,Poor,Nikolov}.~%
\footnote{~%
The author of~\cite{Dombrowski} states that his paper is based on
unpublished lectures of prof.~Willi~Rinow in~1949. See also~\cite[p.~46]{Poor}
where the author claims that the first axiomatical definition of a parallel
transport in the tangent bundle case is given by prof.~W.~Rinow in his
lectures at the Humboldt University in~1949. Some heuristic comments on the
axiomatic approach to parallel transport theory can be found
in~\cite[sec.~2.1]{Gromoll&et_al.} too.%
}
However, a detailed discussion of this topic is out of the scope of the
present work and will be presented elsewhere.

\subsection{Definition and general form}
\label{4-Subsect2.1}

	Let $(E,\pi,B)$ be a complex%
\footnote{%
All of our definitions and results hold also for real vector bundles. Most of
them are valid for vector bundles over more general fields too but this is
inessential for the following.%
}
vector bundle~\cite{Poor,Greub&et_al.-1} with bundle (total) space $E$, base
$B$, projection $\pi\colon E\to B$, and homeomorphic fibres
$\pi^{-1}(x)$, $x\in B$.%
\footnote{~%
When writing $x\in X$, $X$ being a set, we mean ``for all $x$ in $X$'' if
the point $x$ is not specified (fixed, given) and is considered as an
argument or a variable.%
}
Whenever some kind of differentiation in $E$ is
considered, the bundle space $E$ will be required to be a $C^1$
differentiable manifold. The base $B$ is supposed to be a general
topological space in sections~\ref{4-Sect2}--\ref{4-Sect4} and from
Sect~\ref{4-Sect5} onwards is required to be a $C^1$ differentiable manifold.
By $J$ and $\gamma\colon J\to B$ are denoted real interval  and path in $B$,
respectively. The paths considered are generally \emph{not} supposed to be
continuous or differentiable unless their differentiability class is stated
explicitly.

	\begin{Defn}	\label{4-Defn2.1}
	\index{linear transport along paths!definition|defined}
	A \emph{linear transport along paths} in the bundle $(E,\pi,B)$ is a
map $L$ assigning to every path $\gamma$ a map $L^\gamma$,
\emph{transport along} $\gamma$, such that
$L^\gamma\colon (s,t)\mapsto L^\gamma_{s\to t}$ where the map
	\begin{equation}	\label{4-2.1}
L^\gamma_{s\to t} \colon  \pi^{-1}(\gamma(s)) \to \pi^{-1}(\gamma(t))
	\qquad s,t\in J,
	\end{equation}
called \emph{transport along $\gamma$ from $s$ to} $t$, has the properties:
	\begin{alignat}{2}	\label{4-2.2}
L^\gamma_{s\to t}\circ L^\gamma_{r\to s} &=
			L^\gamma_{r\to t},&\qquad  r,s,t&\in J, \\
L^\gamma_{s\to s} &= \id_{\pi^{-1}(\gamma(s))}, & s&\in J,	\label{4-2.3}
\\
L^\gamma_{s\to t}(\lambda u + \mu v) 				\label{4-2.4}
  &= \lambda L^\gamma_{s\to t}u + \mu L^\gamma_{s\to t}v,
	& \lambda,\mu &\in \mathbb{C},\quad u,v\in{\pi^{-1}(\gamma(s))},
	\end{alignat}
where  $\circ$ denotes composition of maps and $\id_X$ is the identity map of
a set $X$.
	\end{Defn}

	\begin{Rem}     \label{4-Rem2.1}
	Equations~\eref{4-2.2} and~\eref{4-2.3} mean that $L$ is a
\emph{transport along paths}
in the bundle $(E,\pi,B)$, which may be an arbitrary topological bundle,
not only a vector one in the general
case~\cite[definition~2.1]{bp-TP-general},~%
\footnote{~%
The definition of a connection in a \emph{topological} bundle $(E,\pi,B)$
in~\cite[ch.~IV, sec.~B.3]{Teleman} is, in fact, an axiomatic definition of a
parallel transport. If we neglect the continuity condition in this
definition, it defines a connection in $(E,\pi,B)$ as a mapping
 $C\colon (\gamma,q)\mapsto C(\gamma,q)$ assigning to any continuous
path $\gamma\colon[0,1]\to B$ and a point $q\in\pi^{-1}(\gamma(0))$ a path
 $C(\gamma,q)\colon[0,1]\to E$ such that $C(\gamma,q)|_0=q$ and
 $\pi\circ C(\gamma,q)=\gamma$. If $I$ is a transport along paths in
$(E,\pi,B)$, then
\(
C\colon (\gamma,q)\mapsto C(\gamma,q)\colon t\mapsto
C(\gamma,q)|_t = I_{0\to t}^{\gamma} (q)
\)
defines a connection $C$ in $(E,\pi,B)$ in the sense mentioned. Moreover, if
this definition is broadened by replacing $[0,1]$ with an arbitrary and not
fixed closed interval $[a,b]$, with $a,b\in\field[R]$ and $a\le b$, then the
converse is also true, \ie
$C(\gamma,q)|_t = I_{a\to t}^{\gamma} (q)$, $t\in[a,b]$, for some transport
$I$. However, the proof of this statement is not trivial.%
}
while~\eref{4-2.4} specifies
that it is \emph{linear}~\cite[equation~(2.8)]{bp-TP-general}. In the present
work only linear transports will be explored.
	\end{Rem}

	\begin{Rem}	\label{4-Rem2.2}
	Definition~\ref{4-Defn2.1} is a generalization of the concept of
`linear connection' given, e.g., in~\cite[sect.~1.2]{Dombrowski} (see
especially~\cite[p.~138, axiom~(L$_1$)]{Dombrowski}) which practically
defines the covariant derivative in terms of linear transports along
paths (see~\eref{4-2.30} below which is equivalent
to~\cite[p.~138, axiom~(L$_3$)]{Dombrowski}). Our definition is much weaker;
\eg  we completely drop~\cite[p.~138, axiom~(L$_3$)]{Dombrowski} and use, if
required, weaker smoothness conditions. An excellent introduction to the
theory of vector bundles and the parallel transports in them can be found in
the book~\cite{Poor}. In particular, in this reference is proved the
equivalence of the concepts parallel transport, connection and covariant
derivative operator in vector bundles (as defined there). Analogous results
concerning linear transports along paths will be presented below. The detailed
comparison of definition~\ref{4-Defn2.1} with analogous ones in the literature
is not a subject of this work and will be given elsewhere (see,
e.g.,~\cite{bp-TP-parallelT}).
	\end{Rem}

	From~\eref{4-2.2} and~\eref{4-2.3}, we get that
$L^\gamma_{s\to t}$ are invertible mappings and
	\begin{equation}	\label{4-2.5}
\left(L^\gamma_{s\to t}\right)^{-1} = L^\gamma_{t\to s},
	\qquad s,t\in J.
	\end{equation}
Hence the linear transports along paths are in fact linear isomorphisms of the
fibres over the path along which they act.

	The following two propositions establish the general structure of
linear transports along paths.%
\footnote{%
Particular examples of proposition~\ref{4-Prop2.1} are known for parallel
transports in vector bundles. For instance, proposition~1
in~\cite[p.~240]{Pham-Mau-Quan} realizes it for parallel transport in a
bundle associated to a principal one and induced by a connection in the
latter case; see also the proof of the lemma in the proof of proposition~1.1
in~\cite[chapter~III, \S~1]{K&N-1}, where a similar result is obtained
implicitly.%
}

	\begin{Prop}	\label{4-Prop2.1}
	\index{linear transport along paths! structure of|(}
	A map~\eref{4-2.1} is a linear transport along $\gamma$ from $s$ to $t$
for every $s,t\in J$ if and only if there exist a vector space $V$,
isomorphic with $\pi^{-1}(x)$ for all $x\in B$, and a family
$\{F(s;\gamma)\colon \pi^{-1}(\gamma(s))\to V,\ s\in J\}$ of linear
isomorphisms such that
	\begin{equation}	\label{4-2.6}
L_{s\to t}^{\gamma} =
	F^{-1}(t;\gamma) \circ  F(s;\gamma),\qquad s,t\in J.
	\end{equation}
	\end{Prop}

	\begin{Proof}
	If~\eref{4-2.1} is a linear transport along $\gamma$ from $s$ to $t$,
then fixing some $s_0\in J$ and using~\eref{4-2.3} and~\eref{4-2.5}, we get
\(
L_{s\to t}^{\gamma} = L_{s_0\to t}^{\gamma} \circ L_{s\to s_0}^{\gamma}
	= \bigl(L_{t\to s_0}^{\gamma}\bigr)^{-1} \circ L_{s\to s_0}^{\gamma}.
\)
So~\eref{4-2.6} holds for $V=\pi^{-1}(\gamma(s_0))$ and
$F(s;\gamma)=L_{s\to s_0}^{\gamma}$. Conversely, if~\eref{4-2.6} is valid for
some linear isomorphisms $F(s;\gamma)$, then a straightforward calculation
shows that it converts~\eref{4-2.2} and~\eref{4-2.3} into identities
and~\eref{4-2.4} holds due to the linearity of $F(s;\gamma)$.
	\end{Proof}

	\begin{Prop}	\label{4-Prop2.2}
	Let a representation~\eref{4-2.6} for a vector space $V$ and some linear
isomorphisms $F(s;\gamma)\colon \pi^{-1}(\gamma(s))\to V,\ s\in J$, be given
for a linear transport along paths in the vector bundle $(E,\pi,B)$. For a
vector space $\lindex[V]{}{\star}$, there exist linear isomorphisms
\(
\lindex[\mspace{-2mu}F]{}{\star}(s;\gamma)\colon \pi^{-1}(\gamma(s))\to
	\lindex[V]{}{\star},
\)
$ s\in J,$ for which
	\begin{equation}	\label{4-2.6'}
L_{s\to t}^{\gamma} =
	\lindex[\mspace{-2mu}F]{}{\star}^{-1}(t;\gamma) \circ
	\lindex[\mspace{-2mu}F]{}{\star}(s;\gamma),\qquad s,t\in J,
	\end{equation}
iff there exists a linear isomorphism
$D(\gamma)\colon V\to\lindex[V]{}{\star}$ such that
	\begin{equation}	\label{4-2.7}
\lindex[\mspace{-2mu}F]{}{\star}(s;\gamma) = D(\gamma)\circ F(s;\gamma),
				\qquad s\in J.
	\end{equation}
	\end{Prop}
\index{linear transport along paths! structure of|)}
%	\vspace{-1.75ex}

	\begin{Proof}
	If equation~\eref{4-2.7} holds, the substitution of
\(
F(s;\gamma) = D^{-1}(\gamma)\circ \lindex[\mspace{-2mu}F]{}{\star}(s;\gamma)
\)
into~\eref{4-2.6} yields~\eref{4-2.6'}. Vice versa, if~\eref{4-2.6'} is valid,
then from its comparison with~\eref{4-2.6} follows that
\(
D(\gamma) = \lindex[\mspace{-2mu}F]{}{\star}(t;\gamma)
				\circ \bigl(F(t;\gamma)\bigr)^{-1}
	  = \lindex[\mspace{-2mu}F]{}{\star}(s;\gamma)
				\circ \bigl(F(s;\gamma)\bigr)^{-1}
\)
is the required (independent of $s,t\in J$) isomorphism.
	\end{Proof}

	Starting from this point, we shall investigate further only the
finite\ndash dimensional case,
 $\dim\pi^{-1}(x)=\dim\pi^{-1}(y)<\infty$ for all $x,y\in B$. In this way
we shall avoid a great number of specific problems arising when the fibres
have infinite dimension (see, e.g.,~\cite{Cooke-infty-mat} for details). A
lot of our results are valid, possibly \emph{mutatis mutandis},  in the
infinite\ndash dimensional treatment too. One way for transferring results
from finite to infinite dimensional spaces is the direct limit from the first
to the second ones. Then, for instance, if the bundle's dimension is
countably or uncountably infinite, the corresponding sums must be replaced by
series or integrals whose convergence, however, requires special
exploration~\cite{Cooke-infty-mat}. Linear transports along paths in
infinite\ndash dimensional vector bundles naturally arise, e.g., in the fibre
bundle formulation of quantum
mechanics~\cite{bp-BQM-introduction+transport,bp-BQM-equations+observables,
	bp-BQM-pictures+integrals,bp-BQM-mixed_states+curvature,
	bp-BQM-interpretation+discussion}.
Generally, there are many difficulties with the infinite\ndash dimensional
problem which deserves a separate investigation.

\subsection{Representations in frames along paths}
\label{4-Subsect2.2}

	Now we shall look locally at linear transports along paths.

	Let $\{e_i(s;\gamma)\}$ be a basis in $\pi^{-1}(\gamma(s))$,
$s\in J$.%
\footnote{%
Here and henceforth the Latin indices run from 1 to
$\dim \pi^{-1}(x),\ x\in B$. We also assume the usual summation rule on
indices repeated on different levels.%
% If the dimension of the bundle is
% countable (resp.\ uncountable) infinity, the corresponding sums must be
% replace by series (resp.\ integrals) whose  convergence is a separate
% problem.%
}
So, along $\gamma\colon J\to B$ we have a set $\{e_i\}$ of bases on
$\pi^{-1}(\gamma(J))$. The dependence of $e_i(s;\gamma)$ on $s$ is
inessential if we are interested only in the \emph{algebraic} properties of
the linear transports along path; this will be the case through the proof of
proposition~\ref{4-Prop2.5}. Starting with two paragraphs before
definition~\ref{4-Defn2.2}, the mapping $s\mapsto e_i(s;\gamma)$ will be
required to be of class $C^1$ as some kind of differentiation of liftings of
paths will be considered.%
\footnote{~%
The mapping $\gamma\mapsto e_i(\cdot,\gamma)$ is, obviously, a lifting of
paths.%
}

	The \emph{matrix}%
\index{matrix!of linear transport along paths|defined[\ff{}]}%
\index{linear transport along paths!matrix of|defined[\ff{}]}
 $\Mat{L}(t,s;\gamma):=\bigl[\Sprindex[L]{j}{i}(t,s;\gamma)\bigr]$
\emph{(along $\gamma$ at $(s,t)$ in $\{e_i\}$) of a linear transport} $L$
along $\gamma$ from $s$ to $t$ is defined via the expansion%
\footnote{%
Notice the different positions of the arguments $s$ and $t$ in
$L_{s\to t}^{\gamma}$ and in $\Mat{L}(t,s;\gamma)$.%
}
	\begin{equation}	\label{4-2.8}
L_{s\to t}^{\gamma} \bigl(e_i(s;\gamma)\bigr)
		=:\Sprindex[L]{i}{j}(t,s;\gamma) e_j(t;\gamma)
		\qquad s,t\in J.
	\end{equation}

	We call $\Mat{L}\colon (t,s;\gamma)\to \Mat{L}(t,s;\gamma)$ the
\emph{matrix (function)} of $L$; respectively $\Sprindex[L]{i}{j}$ are its
\emph{matrix elements}%
\index{linear transport along paths!matrix elements of|defined}
 or \emph{components}%
\index{linear transport along paths!components of|defined}%
\index{components!of linear transport along paths|defined}
 in the given field of bases.

	It is almost evident that
	\begin{equation}	\label{4-2.9}
\Sprindex[L]{i}{j}(t,s;\gamma)e_j(t;\gamma)\otimes e^i(s;\gamma)
\in \pi^{-1}(\gamma(t))\otimes\bigl(\pi^{-1}(\gamma(s))\bigr)^\ast
	\end{equation}
where $\otimes$ is the tensor product sign, the asterisk ($\ast$) denotes
dual object, and $e^i(s;\gamma):=(e_i(s;\gamma))^\ast$. Hence the change of
the bases
\(
  \{ e_i(s;\gamma) \} \mapsto
	\{ e_i^\prime(s;\gamma)	:= A_{i}^{j}(s;\gamma)e_j(s;\gamma) \}
\)
by means of a non\ndash degenerate matrix
$A(s;\gamma):=\bigl[A_{i}^{j}(s;\gamma)\bigr]$ implies%
% \footnote{%
% Here and below we suppose the existence of (two-sided) inverse matrices as
% well as associative matrix products of the matrices we are writing. This is
% trivial in the finite\ndash dimensional case, investigated here, but when
% $\dim\pi^{-1}(x)$ is (countable or uncountable) infinity a lot of
% complications may arise. For details concerning infinite matrices
% see~\cite{Cooke-infty-mat}.%
% }
	\begin{gather}	\label{4-2.10}
\Mat{L}(t,s;\gamma)\mapsto\Mat{L}^\prime(t,s;\gamma)
  = A^{-1}(t;\gamma) \Mat{L}(t,s;\gamma) A(s;\gamma) \displaybreak[1]\\
\intertext{or in component form}
		\tag{\protect\ref{4-2.10}$^\prime$}	\label{4-2.10'}
\Sprindex[L]{i}{\prime\mspace{0.92mu} j}(t,s;\gamma)
   = \bigl(A^{-1}(t;\gamma)\bigr)_{k}^{j}
     \Sprindex[L]{l}{k}(t,s;\gamma) A_{i}^{l}(s;\gamma).
	\end{gather}
Evidently, for $u=u^i(s;\gamma)e_i(s;\gamma)\in\pi^{-1}(\gamma(s))$,
due to~\eref{4-2.4}, we have
	\begin{equation}	\label{4-2.11}
L_{s\to t}^{\gamma} u
   =\bigl( \Sprindex[L]{i}{j}(t,s;\gamma) u^i(s;\gamma) \bigr) e_j(t;\gamma).
	\end{equation}
In terms of the matrix $\Mat{L}$ of $L$, the basic
equations~\eref{4-2.2} and~\eref{4-2.3} read respectively
	\begin{alignat}{2}
	\label{4-2.12}
\Mat{L}(t,s;\gamma) \Mat{L}(s,r;\gamma) &= \Mat{L}(t,r;\gamma)
						&\qquad r,s,t&\in J, \\
	\label{4-2.13}
\Mat{L}(s,s;\gamma) &= \openone	&s&\in J
	\end{alignat}
with $\openone$ being the identity (unit) matrix of corresponding size. From
these equalities immediately follows that $\Mat{L}$ is always non\ndash
degenerate.

	\begin{Prop}	\label{4-Prop2.3}
	A linear map~\eref{4-2.1} is a linear transport along $\gamma$ from
$s$ to $t$ iff its matrix, defined via~\eref{4-2.8}, satisfies~\eref{4-2.12}
and~\eref{4-2.13}.
	\end{Prop}

	\begin{Proof}
	The necessity was already proved. The sufficiency is trivial: a simple
checking proves that~\eref{4-2.12} and~\eref{4-2.13} convert
respectively~\eref{4-2.2} and~\eref{4-2.3} into identities.
	\end{Proof}

	\begin{Prop}	\label{4-Prop2.4}
	A non\ndash degenerate matrix\ndash valued function
$\Mat{L}\colon (t,s;\gamma)\mapsto\Mat{L}(t,s;\gamma)$
is a matrix of some linear transport along paths $L$ (in a given field
$\{e_i\}$ of bases along $\gamma$) iff
	\begin{equation}	\label{4-2.14}
\Mat{L}(t,s;\gamma) = \Mat{F}^{-1}(t;\gamma) \Mat{F}(s;\gamma)
	\end{equation}
where $\Mat{F}\colon (t;\gamma)\mapsto\Mat{F}(t;\gamma)$ is a non\ndash degenerate
matrix\nobreakdash-valued function.
	\end{Prop}

	\begin{Proof}
	This proposition is simply a matrix form of
proposition~\ref{4-Prop2.1}. If $\{f_i\}$ is a basis in $V$ and
 $F(s;\gamma)e_i(s;\gamma)=\Sprindex[F]{i}{j}(s;\gamma)f_j$,
then~\eref{4-2.14} with
$\Mat{F}(s;\gamma) = \bigl[\Sprindex[F]{i}{j}(s;\gamma)\bigr]$
is equivalent to~\eref{4-2.6}.
	\end{Proof}

	\begin{Prop}	\label{4-Prop2.5}
	If the matrix $\Mat{L}$ of a linear transport $L$ along paths has a
representation
	\begin{equation}	\label{4-2.15}
\Mat{L}(t,s;\gamma)
   = \lindex[\mspace{-2mu}\Mat{F}]{}{\star}^{-1}(t;\gamma)
     \lindex[\mspace{-2mu}\Mat{F}]{}{\star}(s;\gamma)
	\end{equation}
for some matrix-valued function
$\lindex[\mspace{-2mu}\Mat{F}]{}{\star}(s;\gamma)$,
then all matrix\nobreakdash-valued functions $\Mat{F}$ representing $\Mat{L}$
via~\eref{4-2.14} are given by
	\begin{equation}	\label{4-2.16}
\Mat{F}(s;\gamma) = \Mat{D}^{-1}(\gamma)
   \lindex[\mspace{-2mu}\Mat{F}]{}{\star}(s;\gamma)
	\end{equation}
where $\Mat{D}(\gamma)$ is a non\ndash degenerate matrix depending only on $\gamma$.
	\end{Prop}

	\begin{Proof}
	In fact, this propositions is a matrix variant of
proposition~\ref{4-Prop2.2};  $\Mat{D}(\gamma)$ is simply the matrix of the map
$D(\gamma)$ in some bases.
	\end{Proof}

	If $\Mat{F}(s;\gamma)$ and $\Mat{F}^\prime(s;\gamma)$ are two
matrix-valued functions, representing the matrix of $L$ via~\eref{4-2.14}
in two bases $\{e_i\}$ and $\{e_i^\prime\}$ respectively, then, as a
consequence of~\eref{4-2.10}, the relation
	\begin{equation}	\label{4-2.16-1}
\Mat{F}^\prime(s;\gamma) = C(\gamma) \Mat{F}(s;\gamma) A(s;\gamma)
	\end{equation}
holds for some non\ndash degenerate matrix\nobreakdash-valued function $C$ of
$\gamma$.

\subsection{Linear transports and derivations along paths}
\label{4-Subsect2.3}

	Below we want to consider some properties of the linear transports
along paths connected with their `differentiability'; in particular, we shall
establish a bijective correspondence between them and the derivations along
paths. For the purpose is required a smooth, of class at least $C^1$,
transition from fibre to fibre when moving along a path in the base.
Rigorously this is achieved by exploring transports in bundles whose
\emph{bundle space is a $C^1$ differentiable manifold} which will be supposed
from now on.

	Let $(E,\pi,B)$ be a vector bundle whose bundle space  $E$ is a $C^1$
differentiable manifold. A linear transport $L^\gamma$ along
$\gamma\colon J\to B$ is called \emph{differentiable of class} $C^k$,
$k=0,1$, or simply $C^k$ transport, if for arbitrary $s\in J$ and
$u\in\pi^{-1}(\gamma(s))$, the path
$\overline{\gamma}_{s;u} \colon J \to E$ with
 $\overline{\gamma}_{s;u}(t) := L^\gamma_{s\to t} u \in \pi^{-1}(\gamma(t))$,
 $t\in J$, is a $C^k$ mapping in the bundle space $E$.%
\footnote{%
If $E$ is of class $C^r$ with $r=0,1,\ldots,\infty,\omega$, we can define in
an evident way a $C^k$ transport for every $k\le r$.%
}
If a $C^k$ linear transport has a representation~\eref{4-2.6}, the mapping
$s\mapsto F(s;\gamma)$ is of class $C^k$. So, the transport $L^\gamma$  is of
class $C^k$ iff $L^\gamma_{s\to t}$ has $C^k$ dependence on  $s$ and $t$
simultaneously. If $\{e_i(\cdot;\gamma)\}$ is a $C^k$ frame along $\gamma$,
\ie $\{e_i(s;\gamma)\}$ is a basis in $\pi^{-1}(\gamma(s))$ and the mapping
$s\mapsto e_i(s;\gamma)$ is of class $C^k$ for all $i$, from~\eref{4-2.11}
follows that $L^\gamma$ is of class $C^k$ iff its matrix $\Mat{L}(t,s;\gamma)$
has $C^k$ dependence on $s$ and $t$.

	Let $E$ be a $C^1$ manifold and $S$ a set of paths in $B$,
$S\subseteq\{\gamma \colon J\to B\}$.
	A transport $L$ along paths in $(E,\pi,B)$, $E$ being $C^r$ manifold,
is said to be of class $C^k$, $k=0,1,\dots,r$, on $S$ if the corresponding
transport $L^\gamma$ along $\gamma$ is of class $C^k$ for all $\gamma\in S$.
	A transport along paths may turn to be of class $C^k$ on some set $S$
of paths in $B$ and not to be of class $C^k$ on other set $S'$ of paths in $B$.
Below, through section~\ref{4-Sect5}, the set $S$ will not be specialized and
written explicitly; correspondingly, we shall speak simply of $C^k$ transports
implicitly assuming that they are such on some set $S$. Starting from
Sect.~\ref{4-Sect5}, we shall suppose $B$ to be a $C^1$ manifold and the set
$S$ to be the one of $C^1$ paths in $B$. Further we consider only $C^1$ linear
transports along paths whose matrices will be referred to smooth frames along
paths.

	Now we want to define what a derivation along paths is (see
definition~\ref{4-Defn2.2} below). For this end we will need some preliminary
material.

	A \emph{lifting} (or lift)%
\footnote{%
For detail see, e.g.,~\cite{Sze-Tsen}.%
}
(in $(E,\pi,B)$) of $g\colon X\to B$, $X$ being a
set, is a map $\overline{g}\colon X\to E$ such that $\pi\circ\overline{g}=g$;
in particular, the liftings of the identity  $\id_B$ of $B$ are called
\emph{sections} and their set is
$\Sec(E,\pi,B):=\{\sigma|\sigma\colon B\to E,\ \pi\circ\sigma=\id_B\}$.
Let $\Path(A):=\{\gamma|\gamma\colon J\to A\}$ be the set of paths in a set
$A$ and
\(
\PLift(E,\pi,B) :=
        \{\lambda|\lambda \colon \Path(B)\to\Path(E),\
        (\pi\circ\lambda)(\gamma) = \gamma
   \text{ for }\gamma\in\Path(B) \}
\)
be the set of liftings of paths from $B$ to $E$.%
\footnote{%
Every linear transport $L$ along paths provides a lifting of paths: for every
$\gamma\colon J\to B$ fix some $s\in J$ and $u\in\pi^{-1}(\gamma(s))$, the
mapping $\gamma\mapsto \overline{\gamma}_{s;u}$ with
$\overline{\gamma}_{s;u}(t):=L_{s\to t}^{\gamma}u$, $t\in J$ is a
lifting of paths from $B$ to $E$.%
}
The set $\PLift(E,\pi,B)$
is:
(i) A natural $\mathbb{C}$\ndash vector space if we put
$ (a\lambda+b\mu)\colon\gamma\mapsto a\lambda_\gamma + b\mu_\gamma $
for $a,b\in\mathbb{C}$, $\lambda,\mu\in\PLift(E,\pi,B)$, and
$\gamma\in\Path(B)$,
where, for brevity, we write $\lambda_\gamma$ for $\lambda(\gamma)$,
$\lambda\colon\gamma \mapsto \lambda_\gamma$;
(ii) A natural left module with respect to complex functions on $B$: if
$f,g\colon B\to \mathbb{C}$, we define
 $(f\lambda+g\mu)\colon\gamma\mapsto(f\lambda)_\gamma+(g\mu)_\gamma$
with $(f\lambda)_\gamma(s):=f(\gamma(s))\lambda_\gamma(s)$ for
$\gamma\colon J\to B$ and $s\in J$;
(iii) A left module with respect to the set
\(
\PF(B) := \{ \varphi|\varphi\colon\gamma\mapsto\varphi_\gamma, \
\gamma\colon J\to B,\ \varphi_\gamma\colon J\to \mathbb{C} \}
\)
of functions along paths in the base $B$:
for $\varphi,\psi\in\PF(B)$, we set
\(
(\varphi\lambda+\psi\mu) \colon\gamma\mapsto
	(\varphi\lambda)_\gamma + (\psi\mu)_\gamma
\)
where
\(
(\varphi\lambda)_\gamma(s)
  := (\varphi_\gamma\lambda_\gamma)(s)
  := \varphi_\gamma(s)\lambda_\gamma(s).
\)

	If we consider $\PLift(E,\pi,B)$ as a $\mathbb{C}$-vector space, its
dimension is equal to infinity. If we regard $\PLift(E,\pi,B)$ as a left
$\PF(B)$\ndash module, its rank is equal to the dimension of $(E,\pi,B)$ (\ie
to the dimension of the fibre(s) of $(E,\pi,B)$). In the last case a basis in
$\PLift(E,\pi,B)$ can be constructed as follows.

	For every path $\gamma\colon J\to B$ and $s\in J$, choose a basis
$\{e_i(s;\gamma)\}$ in the fibre $\pi^{-1}(\gamma(s))$; if the total space
$E$ is a $C^1$ manifold, we suppose $e_i(s;\gamma)$ to have a $C^1$
dependence on $s$. Define liftings along paths $e_i\in\PLift(E,\pi,B)$ by
 $e_i\colon\gamma\mapsto e_i|_\gamma:=e_i(\cdot;\gamma)$, \ie
 $e_i|_\gamma\colon s\mapsto e_i|_\gamma(s):=e_i(s;\gamma)$.
The set $\{e_i\}$ is a basis in $\PLift(E,\pi,B)$, \ie for every
$\lambda\in\PLift(E,\pi,B)$ there are $\lambda^i\in\PF(B)$ such that
$\lambda=\lambda^ie_i$ and $\{e_i\}$ are $\PF(B)$\ndash linearly independent.
Actually, for any path $\gamma\colon J\to B$ and number $s\in J$, we have
$\lambda_\gamma(s)\in\pi^{-1}(\gamma(s))$, so there exists numbers
$\lambda_\gamma^i(s)\in\mathbb{C}$ such that
$\lambda_\gamma(s)=\lambda_\gamma^i(s) e_i(s;\gamma)$.
Defining $\lambda^i\in\PF(B)$ by
$\lambda^i\colon\gamma\mapsto\lambda_\gamma^i$ with
$\lambda_\gamma^i\colon s\mapsto\lambda_\gamma^i(s)$,
we get $\lambda=\lambda^ie_i$;
if $e_i(\cdot;\gamma)$ is of class $C^1$, so are $\lambda_\gamma^i$.
The $\PF(B)$\ndash linear independence of
$\{e_i\}$ is an evident corollary of the $\mathbb{C}$\ndash linear
independence of $\{e_i(s;\gamma)\}$. As we notice above, if $E$ is  $C^1$
manifold, we choose $e_i$, \ie $e_i|_\gamma$, to be of class  $C^1$ and,
consequently, the components $\lambda^i$, \ie $\lambda_\gamma^i$, will be of
class $C^1$ too.

	Let $(E,\pi,B)$ be a vector bundle whose bundle space $E$ is  $C^1$
manifold. Denote by $\PLift^k(E,\pi,B)$, $k=0,1$, the set of liftings of
paths from $B$ to $E$ such that the lifted paths are $C^k$ paths and by
$\PF^k(B)$, $k=0,1$, the set of $C^k$ functions along paths in $B$, \ie
$\varphi\in\PF^k(B)$ if $\varphi_\gamma$ is of class $C^k$.
	Obviously, not every path in $B$ has a $C^k$ lifting in $E$; for
instance, all liftings of a discontinuous path in $B$ are discontinuous paths
in $E$. The set of paths in $B$ having $C^k$ liftings in $E$ is
\(
\pi\circ\Path^k(E)
:= \{ \pi\circ\overline{\gamma} | \overline{\gamma} \in \Path^k(E) \} ,
\)
with $\Path^k(E)$ being the set of $C^k$ paths in $E$. Therefore, when
talking of $C^k$ liftings in $\PLift^k(E,\pi,B)$, we shall implicitly assume
that they are acting on paths in $\pi\circ\Path^k(E)\subset\Path(B)$. The
discontinuous paths in $B$ are, of course, not in $\pi\circ\Path^k(E)$, so
that they are excluded from the considerations below.

	If $E$ and $B$ are $C^1$ manifolds, we denote by $\Sec^k(E,\pi,B)$
the set of $C^k$ sections of the bundle $(E,\pi,B)$.

	\begin{Defn}	\label{4-Defn2.2}
			\index{derivation along paths|defined}
	A \emph{derivation along paths in} $(E,\pi,B)$ or a
\emph{derivation of liftings of paths in} $(E,\pi,B)$ is a map
	\begin{subequations}	\label{4-2.19}
	\begin{equation}	\label{4-2.19a}
	D\colon\PLift^1(E,\pi,B) \to \PLift^0(E,\pi,B)
	\end{equation}
	\end{subequations}
which is $\mathbb{C}$-linear,
	\begin{subequations}	\label{4-2.24}
	\begin{equation}	\label{4-2.24a}
D(a\lambda+b\mu) = aD(\lambda) + bD(\mu)
	\end{equation}
	\end{subequations}
for $a,b\in\mathbb{C}$ and $\lambda,\mu\in\PLift^1(E,\pi,B)$,
and the mapping
	\begin{equation}
	\tag{\protect\ref{4-2.19}b}	\label{4-2.19b}
D_{s}^{\gamma}\colon \PLift^1(E,\pi,B) \to \pi^{-1}(\gamma(s)),
	\end{equation}
defined via
\(
D_{s}^{\gamma}(\lambda)
 := \bigl( (D(\lambda))(\gamma) \bigr) (s)
  = (D\lambda)_\gamma(s)
\)
and called \emph{derivation along} $\gamma\colon J\to B$ \emph{at} $s\in J$,
satisfies the `Leibnitz rule':
	\begin{equation}
	\tag{\protect\ref{4-2.24}b}	\label{4-2.24b}
D_s^\gamma(f\lambda)
 = \frac{\od f_\gamma(s)}{\od s} \lambda_\gamma(s)
	+ f_\gamma(s) D_s^\gamma(\lambda)
	\end{equation}
for every $f\in \PF^1(B)$. The mapping
	\begin{equation}
	\tag{\protect\ref{4-2.19}c}	\label{4-2.19c}
D^{\gamma}\colon \PLift^1(E,\pi,B) \to \Path\bigl(\pi^{-1}(\gamma(J))\bigr),
	\end{equation}
defined by $D^\gamma(\lambda):=(D(\lambda))|_\gamma=(D\lambda)_\gamma$, is
called a \emph{derivation along} $\gamma$.
	\end{Defn}

	Before continuing with the study of linear transports along
paths, we want to say a few words on the links between sections (along paths)
and liftings of paths.

	The set $\PSec(E,\pi,B)$ of sections along paths of $(E,\pi,B)$
consists of mappings
$\bs{\sigma}\colon\gamma\mapsto\bs{\sigma}_\gamma$ assigning
to every path $\gamma\colon J\to B$ a section
$\bs{\sigma}_\gamma\in\Sec\bigl( (E,\pi,B)|_{\gamma(J)} \bigr)$ of the
bundle restricted to $\gamma(J)$. Every (ordinary) section
$\sigma\in\Sec(E,\pi,B)$ generates a section $\bs{\sigma}$ along paths via
\(
\bs{\sigma}\colon\gamma\mapsto\bs{\sigma}_\gamma:=\sigma|_{\gamma(J)},
\)
 \ie
$\bs{\sigma}_\gamma$ is simply the restriction of $\sigma$ to $\gamma(J)$;
hence $\bs{\sigma}_\alpha=\bs{\sigma}_\gamma$ for every path
$\alpha\colon J_\alpha\to B$ with $\alpha(J_\alpha)=\gamma(J)$. Every
$\bs{\sigma}\in\PSec(E,\pi,B)$ generates a lifting
$\hat{\bs{\sigma}}\in\PLift(E,\pi,B)$ by
\(
\Hat{\bs{\sigma}}\colon \gamma\mapsto \Hat{\bs{\sigma}}_\gamma
 := \bs{\sigma}_\gamma\circ\gamma;
\)
in particular, the lifting $\Hat{\sigma}$ associated to
$\sigma\in\Sec(E,\pi,B)$ is given via
$\Hat{\sigma}_\gamma=\sigma|_{\gamma(J)}\circ\gamma$.

	Every derivation $D$ along paths generates a map
\[
\overline{D} \colon \PSec^1(E,\pi,B)\to \PLift^0(E,\pi,B)
\]
such that, if
 $\bs{\sigma}\in \PSec^1(E,\pi,B)$, then
\(
\overline{D}\colon\bs{\sigma} \mapsto \overline{D}\bs{\sigma}
 = \overline{D}(\bs{\sigma})
\)
where
$\overline{D}\bs{\sigma}\colon\gamma\mapsto \overline{D}^\gamma\bs{\sigma}$
is a lifting of paths defined by
\(
\overline{D}^\gamma\bs{\sigma}\colon s\mapsto
(\overline{D}^\gamma\bs{\sigma})(s):=D_s^\gamma \hat{\bs{\sigma}}
\)
with $\hat{\bs{\sigma}}$ being the lifting generated by $\bs{\sigma}$, \ie
$\gamma\mapsto \hat{\bs{\sigma}}_\gamma:= \bs{\sigma}_\gamma\circ \gamma$.
The mapping $\overline{D}$ may be called a derivation of $C^1$ sections along
paths.
Notice, if $\gamma\colon J\to B$ has intersection points and
$x_0\in\gamma(J)$ is such a point, the map $\gamma(J)\to\pi^{-1}(\gamma(J))$
given by
\(
x \mapsto \{ D_s^\gamma(\bs{\sigma}) | \gamma(s)=x,\ s\in J \},
\)
 $x\in\gamma(J)$, is generally multiple\ndash valued at $x_0$ and,
consequently, is not a section of $(E,\pi,B)|_{\gamma(J)}$.

	If $B$ is a $C^1$ manifold and for some $\gamma\colon J\to B$
there exists a subinterval $J'\subseteq J$ on which the restricted path
$\gamma|J\colon J'\to B$ is without self\ndash intersections, \ie
 $\gamma(s)\not=\gamma(t)$ for $s,t\in J'$ and $s\not=t$, we can define
the derivation along $\gamma$ of the sections over $\gamma(J')$ as a map
	\begin{equation}	\label{4-2.17}
\mathsf{D}^\gamma\colon \Sec^1\bigl((E,\pi,B)|_{\gamma(J')}\bigr) \to
         \Sec^0\bigl((E,\pi,B)|_{\gamma(J')}\bigr)
	\end{equation}
such that
	\begin{equation}	\label{4-2.17*}
(\mathsf{D}^\gamma\sigma)(x) := D_{s}^{\gamma}\hat{\sigma}
   \qquad \text{for $x=\gamma(s)$}
	\end{equation}
where $s\in J'$ is unique for a given $x$ and
$\hat{\sigma}\in\PLift\bigl((E,\pi,B)|_{\gamma(J')}\bigr)$ is given by
$\hat{\sigma}=\sigma|_{\gamma(J')}\circ\gamma|_{J'}$. Generally the
map~\eref{4-2.17} defined by~\eref{4-2.17*} is multiple\nobreakdash-valued at
the points of self\ndash intersections of $\gamma$, if any, as
\(
(\mathsf{D}^{\gamma}\sigma)(x)
                := \{ D_{s}^{\gamma}\hat{\sigma} : s\in J,\ \gamma(s)=x \} .
\)
The so\ndash defined map $\mathsf{D}\colon\gamma\mapsto\mathsf{D}^\gamma$ is
called a \emph{section\ndash derivation along paths.}%
\index{section-derivation along paths|defined}
 As we said, it is single\nobreakdash-valued only along paths
without self\ndash intersections.

	Generally a section along paths or lifting of paths does not define a
(single-valued) section of the bundle as well as to a lifting along paths
there does not correspond some (single\ndash valued) section along paths.
	The last case admits one important special exception, \viz if a
lifting $\lambda$ is such that the lifted path $\lambda_\gamma$ is an `exact
topological copy' of the underlying path $\gamma\colon J\to B$, \ie if there
exist $s,t\in J$, $s\not=t$ for which $\gamma(s)=\gamma(t)$, then
$\lambda_\gamma(s)=\lambda_\gamma(t)$, which means that if $\gamma$ has
intersection points, then the lifting $\lambda_\gamma$ also possesses such
points and they are in the fibres over the corresponding intersection points
of $\gamma$.
	Such a lifting $\lambda$ generates a
section $\overline{\lambda}\in\PSec(E,\pi,B)$ along paths given by
 $\overline{\lambda}\colon\gamma\mapsto\overline{\lambda}_\gamma$ with
 $\overline{\lambda}\colon\gamma(s)\mapsto\lambda_\gamma(s)$. In the general
case, the mapping $\gamma(s)\mapsto\lambda_\gamma(s)$ for a lifting $\lambda$
of paths is multiple\ndash valued at the points of self\ndash intersection of
$\gamma\colon J\to B$, if any; for injective path $\gamma$ this map is a
section of $(E,\pi,B)|_{\gamma(J)}$. Such mappings will be called
multiple\ndash valued sections along paths.

	\begin{Defn}	\label{4-Defn2.3}
\index{derivation along paths!generated by linear transport|defined}%
\index{linear transport along paths!derivation along paths generated by|defined[\ff{}]}
        The \emph{derivation $D$ along paths generated by a $C^1$ linear
transport $L$ along paths} in $(E,\pi,B)$, $E$ being a $C^1$ manifold, is a
map of type~\eref{4-2.19a} such that for every path $\gamma\colon J\to B$, we
have
$D^\gamma\colon\lambda\mapsto(D\lambda)_\gamma$ with
$D^\gamma\lambda \colon s  \mapsto D_{s}^{\gamma}\lambda$, $s\in J$,
where  $D_{s}^{\gamma}$ is a map~\protect\eref{4-2.19b} given via
	\begin{equation}	\label{4-2.18}
D_{s}^{\gamma}(\lambda)
   := \lim_{\varepsilon\to0}\Bigl\{\frac{1}{\varepsilon}\bigl[
       L_{s+\varepsilon\to s}^{\gamma}\lambda_\gamma(s+\varepsilon)
       - \lambda_\gamma(s)\bigr]\Bigr\}
	\end{equation}
for every lifting $\lambda\in\PLift^1(E,\pi,B)$
with $\lambda\colon\gamma\mapsto\lambda_\gamma$.
The mapping $D^\gamma$ (resp.\ $D_{s}^{\gamma}$) will be called a
\emph{derivation along $\gamma$ generated by} $L$ (resp.\ a
\emph{derivation along $\gamma$ at $s$ assigned to} $L$).
	\end{Defn}

	\begin{Rem}	\label{4-Rem2.3}
The operator $D_{s}^{\gamma}$ is an analogue of the covariant derivative
assigned to a linear connection;
cf., e.g.,~\cite[p.~139, equation~(12)]{Dombrowski}.
	\end{Rem}

	\begin{Rem}	\label{4-Rem2.3new}
	Notice, if $\gamma$ has self-intersections and $x_0\in\gamma(J)$ is
such a point, the mapping $x\mapsto\pi^{-1}(x)$, $x\in\gamma(J)$, given by
$x\mapsto \{D_s^\gamma(\lambda)|\gamma(s)=x,\ s\in J\}$ is, generally,
multiple\ndash valued at $x_0$.
	\end{Rem}

	Let $L$ be a linear transport along paths in $(E,\pi,B)$. For every
path $\gamma\colon J\to B$, choose some $s_0\in J$ and
$u_0\in\pi^{-1}(\gamma(s_0))$. The mapping
	\begin{equation}	\label{4-2.20new}
\overline{L} \colon\gamma\mapsto \overline{L}_{s_0,u_0}^{\gamma} ,
 \quad  \overline{L}_{s_0,u_0}^{\gamma} \colon J\to E,
 \quad  \overline{L}_{s_0,u_0}^{\gamma} \colon t\mapsto
	\overline{L}_{s_0,u_0}^{\gamma}(t) := L_{s_0\to t}^{\gamma} u_0
	\end{equation}
is, evidently, a lifting of paths.

	\begin{Defn}	\label{4-Defn2.4}
The lifting of paths $\overline{L}$ from $B$ to $E$ in $(E,\pi,B)$ defined
via~\eref{4-2.20new} is called \emph{lifting (of paths) generated by the
(linear) transport} $L$.
	\end{Defn}

	Equations~\eref{4-2.2} and~\eref{4-2.4}, combined with~\eref{4-2.18},
immediately imply
	\begin{gather}		\label{4-2.20}
D_{t}^{\gamma} (\overline{L}) \equiv 0, \qquad t\in J,
			\\	\label{4-2.21}
D_s^\gamma(a\lambda + b\mu)
  = a D_s^\gamma\lambda + b D_s^\gamma\mu,
		\qquad	a,b\in\mathbb{C},
		\quad   \lambda,\mu\in\PLift^1(E,\pi,B),
	\end{gather}
where $s_0\in J$ and $u(s)=L_{s_0\to s}^{\gamma}u_0$ are fixed. In other
words, equation~\eref{4-2.20} means that the lifting $\overline{L}$ is
constant along every path $\gamma$ with respect to $D$.

	Let $\{e_i(s;\gamma)\}$ be a smooth field of bases along
$\gamma\colon J\to B,\ s\in J$. Combining~\eref{4-2.11} and~\eref{4-2.18},
we find the explicit local action of $D_s^\gamma$:%
\footnote{%
The existence of derivatives like $\od\lambda_\gamma^i(s)/\od s$, viz.\ that
$\lambda_\gamma^i\colon J\to\field$ are  $C^1$ mappings, follows from
$\lambda\in\PLift^1(E,\pi,B)$.%
}
	\begin{equation}	\label{4-2.22}
D_s^\gamma\lambda
   = \biggl[ \frac{\od \lambda_\gamma^i(s)}{\od s}
     + \Sprindex[\Gamma]{j}{i}(s;\gamma)\lambda_\gamma^j(s)
     \biggr]  e_i(s;\gamma) .
	\end{equation}
Here the \emph{(2-index) coefficients}%
\index{linear transport along paths!coefficients of|defined[\ff{}]}%
\index{linear transport along paths!coefficients!2-index|defined}%
\index{coefficients!of linear transport along paths|defined[\ff{}]}%
\index{coefficients!2-index of linear transport along paths|defined}
$\Sprindex[\Gamma]{j}{i}$ of the
linear transport $L$ are defined by
	\begin{equation}	\label{4-2.23}
\Sprindex[\Gamma]{j}{i}(s;\gamma)
   := \frac{\pd\Sprindex[L]{j}{i}(s,t;\gamma)}{\pd t}\bigg|_{t=s}
   = - \frac{\pd\Sprindex[L]{j}{i}(s,t;\gamma)}{\pd s}\bigg|_{t=s}
	\end{equation}
and, evidently, uniquely determine the derivation $D$ generated by $L$.

	A trivial corollary of~\eref{4-2.21} and~\eref{4-2.22} is the
assertion that the derivation along paths generated by a linear transport is
actually a derivation along paths (see definition~\ref{4-Defn2.2}).

	Below, we shall prove that, freely speaking, a linear transport along
path(s) can locally, in a given field of local bases, be described
equivalently by the set of its local coefficients (with the transformation
law~\eref{4-2.26} written below).

	If the transport's matrix $\Mat{L}$ has a
representation~\eref{4-2.14}, from~\eref{4-2.23} we get
	\begin{equation}	\label{4-2.25}
\Mat{\Gamma}(s;\gamma):=\bigl[ \Sprindex[\Gamma]{j}{i}(s;\gamma) \bigr]
   = \genfrac{.}{|}{}{}{\pd\Mat{L}(s,t;\gamma)}{\pd t}_{t=s}
   = \Mat{F}^{-1}(s;\gamma)\frac{\od \Mat{F}(s;\gamma)}{\od s} .
	\end{equation}
From here, \eref{4-2.10}, and~\eref{4-2.13}, we see that the change
$\{e_i\}\to\{e_{i}^{\prime}=A_{i}^{j}e_i\}$ of the bases along a path $\gamma$
with a non\ndash degenerate $C^1$ matrix\ndash valued function
$A(s;\gamma):=\bigl[A_{i}^{j}(s;\gamma)\bigr]$ implies
\[
\Mat{\Gamma}(s;\gamma) = \bigl[ \Sprindex[\Gamma]{j}{i}(s;\gamma) \bigr]
	\mapsto  \Mat{\Gamma}^\prime(s;\gamma)
   = \bigl[ \Sprindex[\Gamma]{j}{\prime\,i}(s;\gamma) \bigr]
\]
with
	\begin{equation}	\label{4-2.26}
\Mat{\Gamma}^\prime(s;\gamma)
  = A^{-1}(s;\gamma) \Mat{\Gamma}(s;\gamma) A(s;\gamma)
    + A^{-1}(s;\gamma)\frac{\od A(s;\gamma)}{\od s}.
	\end{equation}

	\begin{Prop}	\label{4-Prop2.5new}
	Let along every (resp.\ given) path $\gamma\colon J\to B$ be given a
geometrical object  $\Gamma$ whose local components
$\Sprindex[\Gamma]{j}{i}$ in a field of bases $\{e_i\}$ along $\gamma$ change
according to~\eref{4-2.26} with
$\Mat{\Gamma}(s;\gamma) = \bigl[ \Sprindex[\Gamma]{j}{i}(s;\gamma) \bigr]$.
There exists a unique linear transport $L$ along paths (resp.\ along $\gamma$)
the matrix of whose coefficients is exactly $\Mat{\Gamma}(s;\gamma)$ in
$\{e_i\}$ along $\gamma$. Moreover, the matrix of the components of $L$ in
$\{e_i\}$ is
	\begin{equation}	\label{4-2.27}
\Mat{L}(t,s;\gamma)
   =  Y(t,s_0;-\Mat{\Gamma}(\cdot;\gamma))
			Y^{-1}(s,s_0;-\Mat{\Gamma}(\cdot;\gamma)),
\qquad s,t\in J
	\end{equation}
where $s_0\in J$ is arbitrarily fixed and the matrix $Y(s,s_0;Z)$, for a
$C^0$ matrix\ndash valued function $Z\colon s\mapsto Z(s)$, is the unique
solution of the initial\ndash valued problem
	\begin{subequations}	\label{4-2.28}
	\begin{align}	\label{4-2.28a}
\frac{\od Y}{\od s}	&= Z(s)Y, \qquad Y=Y(s,s_0;Z), \quad s\in J,
	\\ \label{4-2.28b}
Y(s_0,s_0;Z)		&= \openone.
	\end{align}
	\end{subequations}
	\end{Prop}

	\begin{Proof}
	At the beginning, we note that the proof of existence and uniqueness
of the solution of~\eref{4-2.28} can be found in~\cite[chapter~IV,
\S~1]{Hartman}.

	Given a linear transport $L$ with a matrix~\eref{4-2.14}.
Suppose its components are exactly $\Sprindex[\Gamma]{j}{i}(s;\gamma)$ in
$\{e_i\}$. Solving~\eref{4-2.25} with respect to $\od\Mat{F}^{-1}/\od s$,
we obtain
\(
\od\Mat{F}^{-1}(s;\gamma)/\od s
= - \Mat{\Gamma}(s;\gamma) \Mat{F}^{-1}(s;\gamma)
\)
and, consequently,
\(
\Mat{F}^{-1}(s;\gamma) = Y(s,s_0;-\Mat{\Gamma}(\cdot;\gamma))
\Mat{F}^{-1}(s_0;\gamma).
\)
So, as a result of~\eref{4-2.14}, the matrix of $L$ is~\eref{4-2.27}. Because
of~\cite[chapter~IV, equation~(1.10)]{Hartman}, the expression
\[
Y(t,s;Z) = Y(t,s_0;Z)Y(s_0,s;Z) = Y(t,s_0;Z)Y^{-1}(s,s_0;Z)
\]
is independent of $s_0$. Besides, as a consequence of~\eref{4-2.26}, the
matrix~\eref{4-2.27} transforms according to~\eref{4-2.10} when the local
bases are changed.
Hence~\eref{4-2.3} holds and, due to~\eref{4-2.11}, the linear map
$L$ with a matrix~\eref{4-2.27} in $\{e_i\}$ is a linear transport along
$\gamma$. In this way we have proved two things: On one hand, a linear map
with a matrix~\eref{4-2.27} in $\{e_i\}$ is a linear transport with local
coefficients $\Sprindex[\Gamma]{j}{i}(s;\gamma)$ in $\{e_i\}$ along $\gamma$
and, on the other hand, any linear transport with local coefficients
$\Sprindex[\Gamma]{j}{i}(s;\gamma)$ in $\{e_i\}$ has a matrix~\eref{4-2.27} in
$\{e_i\}$.\hspace*{1em}
	\end{Proof}

	Now we are ready to prove a fundamental result: \emph{\bfseries there
exists a bijective mapping between the sets of $C^1$-linear transports along
paths and derivations along paths}. \emph{The explicit correspondence between
linear transports along paths and derivations along paths is through the
equality of their local coefficients and components, respectively, in a given
field of bases}. After the proof of this result, we shall illustrate it in a
case of linear connections on a manifold.

	\begin{Prop}	\label{4-Prop2.6}
	A mapping~\eref{4-2.19a} (resp.~\eref{4-2.19c}) is a derivation along
paths (resp.\ along $\gamma$) iff there exists a unique linear transport
along paths (resp.\ along $\gamma$) generating it via~\eref{4-2.18}.
	\end{Prop}

	\begin{Proof}
	Let $\{e_i(s;\gamma)\}$ be a frame along $\gamma$ and
$D$ (resp.\ $D^\gamma$) be a derivation along paths (resp.\ along $\gamma$).
Define the \emph{components}%
\index{components!of derivation along paths|defined}%
\index{derivation along paths!components of|defined}%
\footnote{%
In connection with the theory of normal frames (see Sect.~\ref{4-Sect3} and
further), it is convenient to call $\Sprindex[\Gamma]{j}{i}(s;\gamma)$ also
\emph{(2\nobreakdash-index) coefficients}%
\index{coefficients!2-index of derivation along paths|defined[\nn{}]}
\index{coefficients!of derivation along paths|defined[\nn{}]}
of $D^\gamma$. This is consistent with the fact that
$\Sprindex[\Gamma]{j}{i}$  are coefficients of some linear transport
along paths (\emph{see below}).%
}
$\Sprindex[\Gamma]{j}{i}(s;\gamma)$
of $D^\gamma$ in $\{e_i\}$ by the expansion
	\begin{equation}	\label{4-2.29}
D_s^\gamma \hat{e}_j =: \Sprindex[\Gamma]{j}{i}(s;\gamma) e_i(s;\gamma),
	\end{equation}
where $\hat{e}_i\colon\gamma\mapsto e_i(\cdot;\gamma)$ is a liftings of
paths generated by $e_i$. They uniquely define $D^\gamma$
as~\eref{4-2.24} implies~\eref{4-2.22}.
Besides, it is trivial to verify the transformation
law~\eref{4-2.26} for them. So, by proposition~\ref{4-Prop2.5new}, there is a
unique linear transport along paths (resp.\ along  $\gamma$) with the same
local coefficients.

	Conversely, as we already proved, to any linear transport $L$ along
paths (resp.\ along $\gamma$)  there corresponds a derivation $D^\gamma$
along $\gamma$ given via~\eref{4-2.18} whose components coincide with the
coefficients of $L^\gamma$ and transform according to~\eref{4-2.26}.
	\end{Proof}

	We end this section with two examples, the first of which is
quite important and well known.

	Let $\nabla$ be a linear connection (covariant
derivative)~\cite{K&N-1} on a $C^1$ differentiable manifold $M$ and
$\Sprindex[\Gamma]{jk}{i}(x)$, $i,j,k=1,\ldots,\dim M$, $x\in M$, be its local
coefficients in a field $\{E_i(x)\}$ of bases in the tangent bundle over $M$,
\ie  $\nabla_{E_i}E_j= \Sprindex[\Gamma]{ji}{k}E_k$. If $\gamma$ is a $C^1$
path in $M$, then $\nabla_{\dot\gamma}=\dot\gamma^i\nabla_{E_i}$, $\dot\gamma$
being the vector field tangent to $\gamma$, is a derivation along $\gamma$
(in the bundle tangent to $M$) with local components
	\begin{equation}	\label{4-2.30}
\Sprindex[\Gamma]{j}{i}(s;\gamma)
   = \Sprindex[\Gamma]{jk}{i}(\gamma(s)) \dot\gamma^k(s).
	\end{equation}
It is a simple exercise to verify that
\emph{the unique linear transport along paths corresponding, in accordance
with proposition~\ref{4-Prop2.6}, to the derivation with local components
given by~\eref{4-2.30} is exactly the parallel transport generated via the
initial connection $\nabla$}.

	As a second example, we consider a concrete kind of a linear transports
$L$ in the trivial line bundle $(B\times\field[R],\pr_1,B)$, where
$B$ is a topological space, which in particular can be a $C^0$ manifold,
$\times$ is the Cartesian product sign, and
$\pr_1 \colon B\times\field[R]\to B$ is the projection on $B$. An element of
$B\times\field[R]$ is of the form $u=(b,y)$ for some $b\in B$ and
$y\in\field[R]$ and the fibre over $c\in B$ is
$\pr_1^{-1}(c)=\{c\}\times\field[R]=\{(c,z) : z\in\field[R] \}$;
the linear structure of $\pr_1^{-1}(c)$ is given by
$\lambda_1(c,z_1)+\lambda_2(c,z_2)=(c,\lambda_1z_1+\lambda_2z_2)$
 for $\lambda_1,\lambda_2,z_1,z_2\in\field[R]$.
	The bundle $(B\times\field[R],\pr_1,B)$ admits a global frame field
$\{e_1\}$ consisting of a single section $e_1\in\Sec(B\times\field[R],\pr_1,B)$
such that $e_1 \colon B\ni b\mapsto e_1(b)=(b,1)\in\pr_1^{-1}(b)$.
	For $\gamma \colon J\to B$ and $s,t\in J$, define
\(
L \colon \gamma\mapsto L^\gamma
  \colon (s,t)\mapsto L_{s\to t}^\gamma
  \colon \pr_1^{-1}(\gamma(s))\to \pr_1^{-1}(\gamma(t))
\)
by
    \begin{equation}    \label{4-2.30-1}
L_{s\to t}^\gamma(u)
=\Bigl( \gamma(t) , \frac{f(\gamma(s))}{f(\gamma(t))} y \Bigr)
\text{ for }
u=(\gamma(s),y)\in\pr_1^{-1}(\gamma(s)) ,
    \end{equation}
where $f \colon \gamma(J)\to \field[R]\setminus\{0\}$ is a non-vanishing
function on $\gamma(J)$. The verification of~\eref{4-2.2}-\eref{4-2.4} is
trivial and hence $L$ is a linear transport along paths. Its matrix in the
frame $\{e_1\}$ is
\(
\bs{L}(t,s;\gamma)
=\Sprindex[L]{1}{1}(t,s;\gamma)
=\frac{f(\gamma(s))}{f(\gamma(t))} ,
\)
in conformity with~\eref{4-2.14}.
	If $f\circ\gamma \colon J\to \field[R]\setminus\{0\}$
is of class $C^1$, the single coefficient of $L$ is (see~\eref{4-2.23})
 $\Sprindex[\Gamma]{1}{1}(s;\gamma)=\frac{\od }{\od s}\ln(f(\gamma(s)))$;
however, this coefficient is a useful quantity if $B\times\field[R]$ (and hence
$B$) is a $C^1$ manifold --- see~\eref{4-2.22}. Going some pages ahead (see
Proposition~\ref{4-Prop4.2} and definition~\ref{4-Defn3.4} below), we see that
the transport $L$ satisfies equation~\eref{4-4.1} below and therefore admits
normal frames; in particular the frame $\{f_1\}$ such that (see~\eref{4-4.2}
below)
\[
f_1|_{\gamma(s)}
= L_{s_0\to s}^\gamma \bigl( e_1|_{\gamma(s_0)} \bigr)
= \Bigl( \gamma(s) , \frac{f(\gamma(s_0))}{f(\gamma(s))} \Bigr)
\]
for a fixed $s_0\in J$ and any $s\in J$ is normal along $\gamma$, \ie the
matrix of $L$ in $\{f_1\}$ is the identity matrix (the number one in the
particular case).%
\index{linear transport along paths|)}

\section {Normal frames}
\label{4-Sect3}

	The parallel transport in a Euclidean space $\mathbb{E}^n$ (or in
$\mathbb{R}^n$) has the property that, in Cartesian coordinates, it preserves
the components of the vectors that are transported, changing only their
initial points~\cite{DNF-1}. This evident observation, which can be taken
even as a definition for parallel transport in $\mathbb{E}^n$, is of
fundamental importance when one tries to generalize the situation.

	Let a linear transport $L$ along paths be given in a vector bundle
$(E,\pi,B)$, $U\subseteq B$ be an arbitrary subset in $B$, and
$\gamma\colon J\to U$ be a path in $U$.

	\begin{Defn}	\label{4-Defn3.1}
\index{normal frame!in vector bundles!basic definitions|(}
	A \emph{frame field (of bases)} in $\pi^{-1}(\gamma(J))$ is called
\emph{normal along $\gamma$ for} $L$ if the matrix of $L$ in it is the
identity matrix along the given path $\gamma$.
	\end{Defn}

	\begin{Defn}	\label{4-Defn3.2}
	A \emph{frame field (of bases)} defined on $U$ is
called \emph{normal on $U$ for} $L$ if it is normal along every path
$\gamma\colon J\to U$ in $U$. The \emph{frame} is called \emph{normal for}
$L$ if $U=B$.
\index{normal frame!in vector bundles!basic definitions|)}
	\end{Defn}

	Notice that `normal' refers to a `normal form' as opposed to
orthogonal to tangential.

	In the context of the present work, we pose the following problem.
Given a linear transport along paths, is it possible to find a local
basis or a field of bases (frame) in which its matrix is the identity one?
Below we shall rigorously formulate and investigate this problem.~%
\footnote{~%
The problem for exploring normal frames for linear transports seems to be set
in the present paper for the first time. One studies usually normal frames for
some kinds of derivations which, in particular, can be linear
connections~\cite{bp-NF-D+EP}.%
}
	If frames with this property exist, we call them \emph{normal}%
\index{normal frame!in vector bundles|ff}
(for the transport given). According to~\eref{4-2.11}, the linear transports
do not change vectors' components in such a frame and, conversely, a frame
with the last property is normal. Hence the normal frames are a
straightforward generalization of the Cartesian coordinates in Euclidean
space.%
\footnote{%
According to the argument presented, it is more natural to call
Cartesian the special kind of local bases (or frames) we are talking about.
But, in our opinion and for historical reasons, it is better to use the
already established terminology for linear connections and derivations of the
tensor algebra over a differentiable manifold (see below
and~\cite[appendix~A]{bp-PE-P?} or~\cite{bp-Frames-general}).%
}
Because of this and following the established terminology with respect to
metrics~\cite{K&N-1,Bruhat}, we call \emph{Euclidean}%
\index{linear transport along paths!Euclidean|ff}
a linear transport admitting normal frame(s).

	Since a frame field, for instance on a set $U$, is
actually a basis in the set
\(
\Sec\bigl((E,\pi,B)|_U\bigr)
   =\Sec(\pi^{-1}(U),\pi|_U,U),
\)
we call such a basis \emph{normal} if the corresponding field of bases is
normal on $U$.

	\begin{Defn}	\label{4-Defn3.3}
\index{linear transport along paths!Euclidean!basic definitions|(}
	A \emph{linear transport along paths} (or \emph{along a path $\gamma$})
is called \emph{Euclidean along some (or the given) path} $\gamma$ if it admits
a frame normal along $\gamma$.
	\end{Defn}

	\begin{Defn}	\label{4-Defn3.4}
	A \emph{linear transport along paths} is called \emph{Euclidean} on
$U$ if it admits frame(s) normal on $U$. It is called \emph{Euclidean} if
$U=B$.
\index{linear transport along paths!Euclidean!basic definitions|)}
	\end{Defn}

	We want to note that the name ``Euclidean transport'' is connected
with the fact that if we put $B=\mathbb{R}^n$ and
$\pi^{-1}(x)=T_x(\mathbb{R}^n)$ (the tangent space to $\mathbb{R}^n$ at $x$)
and identify $T_x(\mathbb{R}^n)$ with $\mathbb{R}^n$, then in an orthonormal
frame, \ie in Cartesian coordinates, the Euclidean transport coincides with
the standard parallel transport in $\mathbb{R}^n$ (leaving the vectors'
components unchanged).

	Euclidean transports exit always in a case of a trivial bundle
$(B\times V,\pr_1,B)$, with $V$ being a vector space and
$\pr_1 \colon B\times V\to B$ being the projection on $B$; \cf the last
example at the end of section~\ref{4-Subsect2.3}. For instance, the
mapping $L_{s\to t}^\gamma(\gamma(s),v)=(\gamma(t),v)$, for $v\in V$, defines a
Euclidean transport which is similar to the parallel one in $\field[R]^n$.
Indeed, if $\{f_i:i=1,\dots,,\dim V\}$ is a basis of $V$ and $v=v^if_i$, then
$e_i \colon p\mapsto e_i|_p:=(p,f_i)$, $p\in B$, is a (global) frame on $B$ if
we put $v^ie_i|_p=(p,v^if_i)=(p,v)$ and therefore
 $L_{s\to t}^\gamma(e_i|_{\gamma(s)})=e_i|_{\gamma(t)}$, which means that
\(
L \colon \gamma\mapsto L^\gamma
   \colon(s,t) \mapsto L_{s\to t}^\gamma
\)
is a Euclidean transport and $\{e_i\}$ is a normal frame for it (see
corollary~\ref{4-Cor3.1*} below).

	Below we present some general results concerning normal frames
leaving the problem of their existence for the next section.

% 	The importance of the normal frames is established by the following
% result.

	\begin{Prop}	\label{4-Prop3.1}
\index{normal frame!in vector bundles!general properties|(}
	The following statements are equivalent in a given frame $\{e_i\}$
over $U\subseteq B$:

\begin{subequations}	\label{4-3.1}
\indent
\textup{\textbf{\hphantom{v}(i)}}
The matrix of $L$ is the identity matrix on $U$, \ie along every
path $\gamma$ in $U$
	\begin{equation}	\label{4-3.1a}
\Mat{L}(t,s;\gamma) = \openone .
	\end{equation}

\textup{\textbf{\hphantom{i}(ii)}}
The matrix of $L$ along every $\gamma\colon J\to U$
depends only on $\gamma$, \ie it is independent of the points at which it is
calculated:
	\begin{equation}	\label{4-3.1b}
\Mat{L}(t,s;\gamma) = C(\gamma)
	\end{equation}
where $C$ is a matrix-valued function of $\gamma$.

\textup{\textbf{\hphantom{}(iii)}}
If $E$ is a $C^1$ manifold, the coefficients
$\Sprindex[\Gamma]{j}{i}(s;\gamma)$ of $L$ vanish on $U$, \ie along every
path $\gamma$ in $U$
	\begin{equation}	\label{4-3.1c}
\Mat{\Gamma}(s;\gamma) = 0 .
	\end{equation}

\textup{\textbf{\hphantom{}(iv)}}
The explicit local action of the derivation $D$ along paths generated by $L$
reduces on $U$ to differentiation of the components of the liftings with
respect to the path's parameter if the path lies entirely in $U$:
	\begin{equation}	\label{4-3.1d}
D_{s}^{\gamma}\lambda
		= \frac{\od\lambda_\gamma^i(s)}{\od s} \, e_i(s;\gamma)
	\end{equation}
where $\lambda=\lambda^ie_i\in\PLift^1\bigl((E,\pi,B)|_U\bigr)$,
with $E$ being a $C^1$ manifold, and
$\lambda\colon\gamma\mapsto\lambda_\gamma$.

\textup{\textbf{\hphantom{i}(v)}}
The transport $L$ leaves the vectors' components
unchanged along any path in $U$:
	\begin{equation}	\label{4-3.1e}
L_{s\to t}^{\gamma}\bigl(u^i e_i(s;\gamma)\bigr) = u^i e_i(t;\gamma)
	\end{equation}
where $u^i\in\mathbb{C}$.

\textup{\textbf{\hphantom{}(vi)}}
The basic vector fields are $L$\ndash transported
along any path $\gamma\colon J\to U$:
	\begin{equation}	\label{4-3.1f}
L_{s\to t}^{\gamma} \bigl(e_i(s;\gamma)\bigr) =e_i(t;\gamma).
	\end{equation}
	\end{subequations}
	\end{Prop}

	\begin{Proof}
	We have to prove the equivalences
	\begin{multline}	\label{4-3.1new}
\Mat{L}(t,s;\gamma) = C(\gamma)
\iff \Mat{L}(t,s;\gamma) = \openone
\iff    \Mat{\Gamma}(s;\gamma) = 0
\\
\iff  D_{s}^{\gamma}\lambda
	= \frac{\od\lambda_\gamma^i(s)}{\od s} e_i(s;\gamma)
\iff L_{s\to t}^{\gamma}\bigl(u^i e_i(s;\gamma)\bigr) = u^i e_i(t;\gamma)
\\
\iff L_{s\to t}^{\gamma} \bigl(e_i(s;\gamma)\bigr) = e_i(t;\gamma).
	\end{multline}
	If $\Mat{L}(t,s;\gamma)=C(\gamma)$, then, using the
representation~\eref{4-2.14}, we get
 $\Mat{F}(t;\gamma)=\Mat{F}(s;\gamma)C(\gamma)=\Mat{F}(s_0;\gamma)$
for some fixed $s_0\in J$ as $s$ and $t$ are arbitrary, so
\(
\Mat{L}(t,s_0;\gamma)
   = \Mat{F}^{-1}(s_0;\gamma)\Mat{F}(s_0;\gamma) = \openone.
\)
The inverse implication is trivial.
	The second equivalence is a consequence of~\eref{4-2.25}
and~\eref{4-2.14} since $\Mat{\Gamma}=0$ implies
$\Mat{F}(s;\gamma)=\Mat{F}(\gamma)$, while the third one is a corollary
of~\eref{4-2.22}.
	The validity of the last but one equivalence is a consequence of
\(
\Mat{L}(t,s;\gamma) = \openone
\iff L_{s\to t}^{\gamma}\bigl(u^i e_i(s;\gamma)\bigr) = u^i e_i(t;\gamma)
\)
which follows from~\eref{4-2.11}. The last equivalence is a corollary of the
linearity of $L$ and the arbitrariness of $u^i$.
	\end{Proof}

	\begin{Rem}	\label{4-Rem3.1}
	An evident corollary of the last proof is
	\begin{equation}	\label{4-3.2a}
\Mat{L}(t,s;\gamma) = \openone \iff \Mat{F}(s;\gamma) = \Mat{B}(\gamma)
	\end{equation}
with $\Mat{B}$ being a matrix-valued function of the path $\gamma$ only.
According to proposition~\ref{4-Prop2.5}, this dependence is inessential and,
consequently, in a normal frame, we can always choose
representation~\eref{4-2.14} with
	\begin{equation}	\label{4-3.2b}
\Mat{F}(s;\gamma) = \openone.
	\end{equation}
	\end{Rem}

	\begin{Cor}	\label{4-Cor3.1*}
	The equalities~\eref{4-3.1a}\Ndash\eref{4-3.1f} are equivalent and
any one of them express a necessary and sufficient condition for a frame to
be normal for $L$ in $U$. In particular, for $U=\gamma(J)$ they express such
a condition along a fixed path $\gamma$.
	\end{Cor}

	\begin{Proof}
	This result is a direct consequence of definition~\ref{4-Defn3.2} and
proposition~\ref{4-Prop3.1}.
	\end{Proof}

% The old text follows
% 	\begin{Cor}	\label{4-Cor3.1*}
% 	Any one of the equalities~\eref{4-3.1a}\Ndash\eref{4-3.1f}
% express a necessary and sufficient condition for a frame to be normal for
% $L$ in $U$.
% 	\end{Cor}
% 	\begin{Proof}
% 	This result is a direct consequence of definition~\ref{4-Defn3.2} and
% proposition~\ref{4-Prop3.1}.
% 	\end{Proof}
%
% 	\begin{Prop}	\label{4-Prop3.1*}
% 	The equations~\eref{4-3.1a}--\eref{4-3.1f} are equivalent in a given
% frame $\{e_i\}$ along a (fixed) path $\gamma\colon J\to B$.
% 	\end{Prop}
% 	\begin{proof}
% 	This proof is identical with the one of proposition~\ref{4-Prop3.1} for
% $U=\gamma(J)$.
% 	\end{proof}
%
% 	\begin{Cor}	\label{4-Cor3.1**}
% 	A frame is normal along $\gamma$ for $L$ if and only if in that frame
% one (and hence all) of the equalities~\eref{4-3.1a}--\eref{4-3.1f} is (are)
% valid.
% 	\end{Cor}
%
% 	\begin{Proof}
% 	The result follows from definition~\ref{4-Defn3.1} and
% proposition~\ref{4-Prop3.1*}.
% 	\end{Proof}
%
% 	In particular, a frame is normal for $L$ along $\gamma$ iff in the
% frame the coefficients of $L$ vanish along $\gamma$, \ie iff~\eref{4-3.1c}
% holds.

	A lifting of paths $\lambda\in\PLift(E,\pi,B)$ is called
$L$\ndash transported along $\gamma\colon J\to B$, if for every $s,t\in J$ is
fulfilled
\(
\lambda_\gamma(t)=L_{s\to t}^{\gamma}\lambda_\gamma(s)
\)
with
$\lambda\colon\gamma\mapsto\lambda_\gamma$. Hence a frame $\{e_i(s,\gamma)\}$
along $\gamma$ is $L$\ndash transported along $\gamma$ if the basic vectors
$\Hat{e}_1,\dots,\Hat{e}_{\dim B}$, considered as liftings of paths,
\ie $\Hat{e}_i\colon \gamma\mapsto e_i(\cdot;\gamma)$,
are $L$\ndash transported along $\gamma$.

	Therefore a frame is normal for $L$ along $\gamma$ iff it is
$L$\ndash transported along $\gamma$, \ie if, by definition, its basic
vectors $e_i(s;\gamma)$ satisfy~\eref{4-3.1f}. As we shall see below (see
proposition~\ref{4-Prop3.10}), this allows a convenient and useful way for
constructing normal frames, if any.

	For the above reasons, sometimes, it is convenient for the
definition~\ref{4-Defn3.1} to be replaced, equivalently, by the next ones.

	\renewcommand{\theVarDefn}{\protect\ref{4-Defn3.1}$^{\bs{\prime}}$}
	\begin{VarDefn}	\label{4-Defn3.1'}
	If $E$ is a $C^1$ manifold, a \emph{frame (or frame field)}
over $\gamma(J)$ is called
\emph{normal along} $\gamma\colon J\to B$ for a linear transport $L$ along
paths if the coefficients of $L$ along $\gamma$ vanish in it.
	\end{VarDefn}

	\renewcommand{\theVarDefn}%
		     {\protect{\ref{4-Defn3.1}}$^{\bs{\prime}\bs{\prime}}$}
	\begin{VarDefn}	\label{4-Defn3.1''}
	A \emph{frame} over $\gamma(J)$ is called \emph{normal along}
$\gamma\colon J\to B$ for a linear transport $L$ along paths if it is
$L$\ndash transported along $\gamma$.
	\end{VarDefn}
\noindent The last definition of a normal frame is, in a sense, the `most
invariant (basis\ndash free)' one.

	The next proposition describes the class of normal frames, if any,
along a given path.

	\begin{Prop}	\label{4-Prop3.2}
	All frames normal for some linear transport along paths which is
Euclidean along a certain (fixed) path  are connected by linear
transformations whose matrices may depend only on the given path but not on
the point at which the bases are defined.
	\end{Prop}

	\begin{Proof}
	Let $\{e_i\}$ and $\{e_i^\prime:=A_{i}^{j}e_j\}$ be frames normal
along $\gamma\colon J\to B$ for a linear transport $L$ along paths and
$\Mat{L}$ and $\Mat{L}^\prime$ be the matrices of $L$ in them respectively.
As, by definition $\Mat{L}=\Mat{L}^\prime=\openone$, from~\eref{4-2.10}, we
get $A(s;\gamma)=A(t;\gamma)$ for any $s,t\in J$, \ie $A(s;\gamma)$ depends
only on $\gamma$ and not on $s$.

	If $E$ is a $C^1$ manifold and $\Mat{\Gamma}$ and
$\Mat{\Gamma}^\prime$ are the matrices of the coefficients of $L$ in $\{e_i\}$
and $\{e'_i\}$, respectively, by proposition~\ref{4-Prop3.1} we have
$\Mat{\Gamma}=\Mat{\Gamma}^\prime=0$, so the transformation law~\eref{4-2.26}
implies $\od A(s;\gamma)/\od s = 0$,
$A(s;\gamma):=\bigl[A_{i}^{j}\bigr(s;\gamma)]$.
	\end{Proof}

	\begin{Cor}	\label{4-Cor3.1}
	All frames normal for a Euclidean transport along a given path are
obtained from one of them via linear transformations whose matrices may depend
only on the path given but not on the point at which the bases are defined.
	\end{Cor}

	\begin{Proof}
	See proposition~\ref{4-Prop3.2} or its proof.
	\end{Proof}

	The following two results describe the class of all frames normal on an
arbitrary set $U$, if such frames exist.

	\begin{Cor}	\label{4-Cor3.2}
	If a linear transport along paths admits frames normal on a set $U$,
then all of them are connected via linear transformations with constant (on
$U$) matrices.
	\end{Cor}

	\begin{Proof}
	Let $\{e_i\}$ and $\{e_i^\prime:=A_{i}^{j}e_i\}$ be  frames
normal on $U$ and $x\in U$. By proposition~\ref{4-Prop3.2} (see also
definition~\ref{4-Defn3.2}), for any paths $\beta$ and $\gamma$ in $U$ passing
though $x$, we have
 $A(x):=\bigl[A_{i}^{j}\bigr]=\Mat{B}(\beta)=\Mat{B}(\gamma)$ for some
matrix\ndash valued function $\Mat{B}$ on the set of the paths in $U$. Hence
$A(x)=\const$ on $U$, due to the arbitrariness of $\beta$ and $\gamma$.
	\end{Proof}

	\begin{Cor}	\label{4-Cor3.3}
	If a linear transport along paths admits a frame normal on a set $U$,
then all such frames on $U$ for it are obtained from that frame by linear
transformations with constant (on $U$) coefficients.
	\end{Cor}

	\begin{Proof}
	The result immediately follows from corollary~\ref{4-Cor3.2}
	\end{Proof}

	We end this section with a simple but important result which shows
how the normal frames, if any, can be constructed along a given path.

	\begin{Prop}	\label{4-Prop3.10}
	If $L$ is Euclidean transport along $\gamma\colon J\to B$ and
$\{e_{i}^{0}\}$ is a basis in $\pi^{-1}(\gamma(s_0))$ for some $s_0\in J$,
then the frame $\{e_{i}^{}\}$ along $\gamma$ defined by
	\begin{equation}	\label{4-3.10}
e_i(s;\gamma) = L_{s_0\to s}^{\gamma} \bigl( e_{i}^{0} \bigr), \qquad s\in J
	\end{equation}
is normal for $L$ along $\gamma$.
	\end{Prop}

	\begin{Proof}
	Due to~\eref{4-2.2} and~\eref{4-3.10}, the frame $\{e_{i}\}$
satisfies~\eref{4-3.1f} along $\gamma$. Hence, by corollary~\ref{4-Cor3.1*},
it is normal for $L$ along $\gamma$.
	\end{Proof}

	An analogous result on a set $U\subseteq B$ will be presented in the
next section (see below - proposition~\ref{4-Prop4.3}).%
\index{normal frame!in vector bundles!general properties|)}

\section {On the existence of normal frames}
\label{4-Sect4}
\index{normal frame!in vector bundles!existence theorems|(}

	In the previous section there were derived a number of properties of the
normal frames, but the problem of their existence was neglected. This is
the subject of the present section.

	At a given point $x\in B$ the following result is valid.

	\begin{Prop}	\label{4-Prop4.1}
	A linear transport $L^\gamma$ along $\gamma\colon J\to B$ such that
$\gamma(J)=\{x\}$ for a given point $x\in B$ admits normal frame(s) iff it is
the identity mapping of the fibre over $x$, \ie
$L_{s\to t}^{\gamma}=\id_{\pi^{-1}(x)}$ for every $s,t\in J$.
	\end{Prop}

	\begin{Proof}
	The sufficiency is trivial (see definition~\ref{4-Defn2.1}). If
$\{e_i\}$ is normal for $L^\gamma$ (at $x$), then
\(
L_{s\to t}^{\gamma}(u^ie_i|_x)
   = u^i L_{s\to t}^{\gamma} e_i|_x = u^i e_i|_x,
\)
 $u^i\in\mathbb{C}$ due to $\gamma(s)=\gamma(t)=x$ and
proposition~\ref{4-Prop3.1}, point~(iv). Therefore
$L_{s\to t}^{\gamma}=\id_{\pi^{-1}(x)}$.
	\end{Proof}

	Thus, for a degenerate path $\gamma\colon J\to\{x\}\subset B$ for some
$x\in B$, the identity mapping of the fibre over $x$ is the only realization
of a Euclidean transport along paths. Evidently, for such a transport every
basis of that fibre is a frame normal at $x$ for it.

	\begin{Prop}	\label{4-Prop4.2}
	A linear transport $L$ along paths admits frame(s) normal along a
given path $\gamma\colon J\to B$ iff
	\begin{equation}	\label{4-4.1}
L_{s\to t}^{\gamma}=\id_{\pi^{-1}(\gamma(s))}
\qquad
\text{\upshape for every $s,t\in J$ such that $\gamma(s)=\gamma(t)$},
	\end{equation}
i.e., if $\gamma$ contains loops, the $L$-transport along each of them
reduces to the identity mapping of the fibre over the initial/final point of
the transportation.
	\end{Prop}

	\begin{Rem}	\label{4-Rem4.1}
 For $s=t$ the equation~\eref{4-4.1} is identically satisfied due
to~\eref{4-2.3}. But for $s\not=t$, if such $s$ and $t$ exist, this is highly
non\ndash trivial restriction: it means that the result of $L$\ndash
transportation along $\gamma$ of a vector $u\in\pi^{-1}(x_0)$ for some
$x_0\in\gamma(J)$ from $x_0$ to a point $x\in\gamma(J)$ is independent of how
long the vector has `traveled' along $\gamma$ or, more precisely, if
$x_0,x\in\gamma(J)$ are fixed and, for each $y\in\gamma(J)$, $J_y:=\{r\in J :
\gamma(r)=y\}$, then the vector
 $L_{s_0\to s}^\gamma(u)$ is independent of the choice of the points
 $s_0\in J_{x_0}$ and $s\in J_x$ (if some of the sets $J_{x_0}$ and/or $J_x$
contain more than one point). This is trivial if $\gamma$ is without
self\ndash intersections (see~\eref{4-2.2}). If $\gamma$ has self\ndash
intersections, \eg if $\gamma$ intersects itself one time at $\gamma(s)$, \ie
if $\gamma(s)=\gamma(t)$ for some $s,t\in J,\ s\not=t$, then the result of
$L$\ndash transportation of $u\in\pi^{-1}(\gamma(s_0))$ from
$x_0=\gamma(s_0)$ to $x=\gamma(s)=\gamma(t)$ along $\gamma$ is
%
% \footnote{%
% The number of the obtained values at $x=\gamma(s)$ is one plus the number of
% self\ndash intersections of $\gamma$ at $x$ if $\gamma$ has a unique point of
% self\ndash intersection.%
% }
\( u_s=L_{s_0\to s}^{\gamma}u \)
or
\( u_t=L_{s_0\to t}^{\gamma}u \).
We have $u_s=u_t$ iff~\eref{4-4.1} holds. Rewording, if we fix some
$u_0\in\pi^{-1}(\gamma(s_0))$, the bundle\nobreakdash-valued function
$u\colon \gamma(J)\to E$ given by
\(
u\colon \gamma(s)\to
u_s=L_{s_0\to s}^{\gamma}u_0 \in \pi^{-1}(\gamma(s))
\)
for $s\in J$
is single\nobreakdash-valued iff~\eref{4-4.1} is valid.%
\footnote{%
The so-defined map $u$ is a section along $\gamma$ of
$(E,\pi,B)$~\cite{bp-BQM-equations+observables}. Generally it is a
multiple\nobreakdash-valued map (see Sect.~\ref{4-Sect2}).%
}
Notice, since $\pi\circ u_s\equiv\gamma(s)$ (see~\eref{4-2.1}), the map $u$
is (a single\nobreakdash-valued) lifting of $\gamma$ in $E$ through
$u_0$ irrespectively of the validity of~\eref{4-4.1}.

	\emph{Prima facie} the above may be reformulated in terms of the
concept of holonomy in vector bundles~\cite[pp.~51--54]{Poor}. But a rigorous
analysis reveals that this is impossible in the general case without imposing
further restrictions, like equation~\eref{4-4.3} below, on the transports
involved. For instance, without requiring equation~\eref{4-4.3} below to be
valid, one cannot introduce the concept of a holonomy group.
	\end{Rem}

	\begin{Proof}
	If $L$ is Euclidean along $\gamma$, then ~\eref{4-4.1} follows from
equation~\eref{4-3.1e} as it holds for every $u^i\in\mathbb{C}$ in some
normal frame $\{e_i\}$. Conversely, let~\eref{4-4.1} be valid. Put
	\begin{equation}	\label{4-4.2}
e_i|_{\gamma(s)} := L_{s_0\to s}^{\gamma} \bigl( e_{i}^{0} \bigr)
	\end{equation}
where $\{e_{i}^{0}\}$ is a fixed basis in $\pi^{-1}(\gamma(s_0))$ for a fixed
$s_0\in J$. Due to the nondegeneracy of $L$,
$\{e_{i}^{}\}$ is a basis at $\gamma(s)$ for every $s$.
According to~\eref{4-4.1}, the so\ndash defined field of bases $\{e_{i}^{}\}$
along $\gamma$ is single\nobreakdash-valued. By means of~\eref{4-2.2}, we
easily verify that~\eref{4-3.1f} holds for $\{e_{i}^{}\}$. Hence
$\{e_{i}^{}\}$ is normal for $L$ along $\gamma$.
	\end{Proof}

	\begin{Rem}	\label{4-Rem4.2}
	Regardless of the validity of~\eref{4-4.1}, equation~\eref{4-4.2}
defines a field of, generally multiple\nobreakdash-valued, normal frames in
the set of sections along $\gamma$ of $(E,\pi,B)$.
(For details on sections along paths, see Sect.~\ref{4-Sect2}.)

	Such a multi-valued property can be avoid if $\gamma$ is supposed to be
injective ($\Leftrightarrow$ without self-intersections). \emph{Prima facie} one
may think that this solves the multi-valued problem in the general case by
decomposing $\gamma$ into a union of injective paths. However, this is not the
most general situation because a transport along a composition of paths  does
not generally equal to the composition of the transports along its constituent
sub-paths (see equation~\eref{4-4.3}  below); besides, since
equation~\eref{4-4.7} below does not hold generally, the absents of a
natural/canonical definition of composition (product) of paths introduces an
additional indefiniteness.
	\end{Rem}

	\begin{Cor}	\label{4-Cor4.1}
	Every linear transport along paths is Euclidean along every fixed
path without self\ndash intersections.
	\end{Cor}

	\begin{Proof}
	For a path $\gamma\colon J\to B$ without self\ndash intersections, the
equality $\gamma(s)=\gamma(t)$, $s,t\in J$ is equivalent to $s=t$. So,
according to~\eref{4-2.3}, the condition~\eref{4-4.1} is identically satisfied.
	\end{Proof}

	Now we shall establish an important necessary and sufficient condition
for the existence of frames normal on an arbitrary subset $U\subseteq B$.

	\begin{Thm}	\label{4-Thm3.1}
	A linear transport along paths admits frames normal on some set
(resp.\ along a given path) if and only if its action along every path in
this set (resp.\ along the given path) depends only on the initial and final
point of the transportation but not on the particular path connecting these
points. In other words, a transport is Euclidean on $U\subseteq B$ iff it
is path\ndash independent on $U$.
	\end{Thm}

	\begin{Proof}
	Let a linear transport $L$ admit a frame $\{e_{i}\}$ normal in
$U\subseteq B$. By definitions~\ref{4-Defn3.1} and~\ref{4-Defn3.2} and
equation~\eref{4-2.11}, this implies
\(
L_{s\to t}^{\gamma}u^i(\gamma(s)) \bigl( e_i|_{\gamma(s)} \bigr)
   = u^i(\gamma(s)) e_i|_{\gamma(t)}
\)
for $\gamma\colon J\to U$ and $u(x)\in\pi^{-1}(x)$, $x\in B$.
Conversely, let
 $L_{s\to t}^{\gamma} u(\gamma(s))$ depend only on $\gamma(s)$ and
$\gamma(t)$ but not on $\gamma$ and $\{e_i\}$ be a field of bases on $U$
(resp.\ on $\gamma(J)$). Then, due to~\eref{4-2.11}, the matrix $\Mat{L}$ of $L$
in $\{e_i\}$ has the form
$\Mat{L}(t,s;\gamma) = \Mat{B}(\gamma(t),\gamma(s))$ for some
matrix\nobreakdash-valued function $\Mat{B}$ on $U\times U$. Combining this
result with propositions~\ref{4-Prop2.4} and~\ref{4-Prop2.5}, we see that
$\Mat{L}$ admits a representation
	\begin{equation}	\label{4-3.3}
\Mat{L}(t,s;\gamma) = \Mat{F}_{0}^{-1}(\gamma(t)) \Mat{F}_{0}(\gamma(s)),
   \qquad s,t\in J
	\end{equation}
for a non\ndash degenerate matrix-valued function $\Mat{F}_{0}$ on $U$.
At last, putting
\(
e_{i}^{\prime}|_x
   = \Sprindex[\bigl(\Mat{F}_{0}^{-1}(x)\bigr)]{i}{j} e_j|_x,
\)
$ x\in U$,
from~\eref{4-2.10} we obtain that the matrix of $L$ in $\{e_i^\prime\}$ is
$\Mat{L}^\prime(t,s;\gamma)=\openone$, \ie the frame $\{e_i^\prime\}$ is
normal for $L$ on $U$.
	\end{Proof}

	An evident corollary of theorem~\ref{4-Thm3.1} is the following
assertion. Let a linear transport $L$ be Euclidean on $U\subseteq B$ and
$h_a\colon J\to U$, $a\in[0,1]$, be a homotopy of paths passing through two
fixed points $x,y\in U$, \ie $h_a(s_0)=x$ and $h_a(t_0)=y$ for some
$s_0,t_0\in J$ and any $a\in[0,1]$. Then  $L_{s_0\to t_0}^{h_a}$ is
independent of $a\in[0,1]$. In particular, we have
$L_{s_0\to t_0}^{h_a}\big|_{y=x}=\id_{\pi^{-1}(x)}$ owing to
proposition~\ref{4-Prop4.2}.

	Equation~\eref{4-3.3} and the part of the proof of theorem~\ref{4-Thm3.1}
after it are a hint for the formulation of the following result.

	\begin{Thm}	\label{4-Thm3.2}
	A linear transport $L$ along paths in a vector bundle, with $C^1$
manifold as a bundle space, is Euclidean on $U$ (resp.\ along $\gamma$) iff
for some, and hence for every, frame $\{e_i\}$ on $U$ (resp.\ on $\gamma(J)$)
there exists a non\ndash degenerate matrix-valued function $\Mat{F}_0$ on $U$
such that the matrix $\Mat{L}$ of $L$ in $\{e_i\}$ is given by~\eref{4-3.3}
for every $\gamma\colon J\to U$ (resp.\ for the given $\gamma$) or,
equivalently, iff the matrix $\Mat{\Gamma}$ of the coefficients of $L$ in
$\{e_i\}$ is

	\begin{equation}
		\tag{\protect\ref{4-3.3}$^\prime$}	\label{4-3.3'}
\Mat{\Gamma}(s;\gamma)
   = \Mat{F}_{0}^{-1}(\gamma(s)) \frac{\od\Mat{F}_0(\gamma(s))}{\od s} .
	\end{equation}
	\end{Thm}

	\begin{Proof}
	Suppose $L$ is Euclidean. There is a frame $\{e_{i}^{0}\}$ normal for
$L$ on $U$ (resp.\ along $\gamma$). Define a matrix $\Mat{F}_0(x)$ via the
expansion \( e_i|_x
   = \Sprindex[\bigl(\Mat{F}_0(x)\bigr)]{i}{j} e_{j}^{0}\big|_x,
\)
 $x\in U$.
Since, by definition, the matrix of $L$ in $\{e_{i}^{0}\}$ is the
unit (identity) matrix on $U$, the matrix of $L$ in $\{e_{i}\}$ is given
via~\eref{4-3.3} due to~\eref{4-2.10}. Conversely, if~\eref{4-3.3} holds in
$\{e_{i}\}$ on $U$, the frame
\(
\{ e_{i}^{\prime}|_x
= \Sprindex[\bigl(\Mat{F}_0^{-1}(x)\bigr)]{i}{j} e_{j}^{0}\big|_x \}
\)
is normal for $L$ on $U$ (resp.\ along $\gamma$), as we saw at the end of the
proof of theorem~\ref{4-Thm3.1}. The equivalence of~\eref{4-3.3'}
and~\eref{4-3.3} is a consequence of~\eref{4-2.23} (cf.~\eref{4-2.25},
\eref{4-2.26}, and~\eref{4-3.1new}).
	\end{Proof}

	The proof of theorem~\ref{4-Thm3.2} suggest a way for generating
Euclidean transports along paths by `inverting' the definition of normal
frames: take a given field of bases over $U\subseteq B$ and define a linear
transport by requiring its matrix to be unit in the given field of bases.
% More precisely, we have in mind the following. Let $\{e_i\}$ be a fixed
% frame on $U$, $\{e_i^\prime =A_{i}^{j}e_j\}$, with
% $A=\bigl[A_{i}^{j}\bigr]$ being non\ndash degenerate, be arbitrary frame on
% $U$, and $\gamma\colon J\to U$ be a path in $U$. Define a linear
% map~\eref{4-2.1} by its matrix in $\{e_i^\prime\}$ (see~\eref{4-2.11}):
% 	\begin{equation}	\label{4-3.4}
% \Mat{L}(t,s;\gamma) = A(\gamma(t))A^{-1}(\gamma(s)),
%    \qquad A(x) := \bigl[A_{i}^{j}(x)\bigr], \quad x\in U.
% 	\end{equation}
% According to proposition~\ref{4-Prop2.4}, the map
%  $L\colon \gamma\mapsto L^\gamma\colon (s,t)\mapsto L_{s\to t}^{\gamma}$
% is a linear transport along paths. By theorem~\ref{4-Thm3.1}, this transport
% is Euclidean. Moreover, from~\eref{4-2.10}, we see that the matrix of $L$
% in $\{e_i\}$ is identity (unit) on $U$, \ie $\{e_i\}$ is a frame normal for
% $L$ on $U$.
      We call this Euclidean transport \emph{generated by (or assigned to)}%
\index{linear transport along paths!Euclidean!generated by frame}
 the given initial frame, which is normal for it.

	\begin{Prop}	\label{4-Prop3.3}
	All frames normal for a Euclidean transport along paths in $U$
generate one and the same Euclidean transport along paths in $U$ coinciding
with the initial one.
	\end{Prop}
	\begin{proof}
	The result is an almost evident consequence of
% ~\eref{4-3.4}
 the last definition
and corollary~\ref{4-Cor3.3}.
	\end{proof}

	\begin{Prop}	\label{4-Prop3.4}
	Two or more frames on $U$ generate one and the same
Euclidean transport along paths iff they are connected via linear
transformations with constant (on $U$) coefficients.
	\end{Prop}

	\begin{Proof}
	If $\{e_i\}$ and $\{e_i^\prime\}$ generate $L$, then they are normal
for it (proposition~\ref{4-Prop3.3}) and, by corollary~\ref{4-Cor3.2}, they are
connected in the way pointed. The converse is a trivial corollary
of
% ~\eref{4-3.4} for $A(x)=\const$ with $x\in U$.
 the last definition.
	\end{Proof}

	In this way we have established a bijective correspondence between
the set of Euclidean linear transports along paths in $U$ and the class of
sets of frames on $U$ connected by linear transformations with constant
coefficients.

	The comparison of proposition~\ref{4-Prop4.2} with theorem~\ref{4-Thm3.1}
suggests that a transport is Euclidean in $U\subseteq B$ iff~\eref{4-4.1} holds
for every $\gamma\colon J\to U$. But this is not exactly the case. The right
result is the following one.

	\begin{Thm}	\label{4-Thm4.3}
	A linear transport $L$ along paths is Euclidean on a
path\ndash connected set %
% \footnote{%
% A set is path\ndash connected if every two its points can be connected by a
% continuous path lying entirely in it. Sometimes such sets are called
% linearly connected.%
% }
 $U\subseteq B$ iff the next three conditions are valid:
(i) Equation~\eref{4-4.1} holds for every continuous path
$\gamma\colon J\to U$;
(ii) The transport along a product of paths is equal to the composition of
the transports long the paths of the product, \ie
	\begin{equation}	\label{4-4.3}
L^{\gamma_1\gamma_2} = L^{\gamma_2}\circ L^{\gamma_1}
	\end{equation}
where $\gamma_1$ and $\gamma_2$ are paths in $U$ such that the end of
$\gamma_1$ coincides with the beginning of $\gamma_2$ and  $\gamma_1\gamma_2$
is the product of these paths;
(iii) For any subinterval $J'\subseteq J$ the locality condition
	\begin{equation}	\label{4-4.4}
L_{s\to t}^{\gamma|J'} = L_{s\to t}^{\gamma}, \qquad s,t\in J'\subseteq J,
	\end{equation}
with $\gamma|J'$ being the restriction of $\gamma\colon J\to U$ to $J'$,
is valid.
	\end{Thm}

	\begin{Rem}	\label{4-Rem4.3a}
	Here and below we do not present and use a particular definition of
the product of paths. There are slightly different versions of that
definition; for details see~\cite{Viro&Fuks,Sze-Tsen}
or~\cite[sect.~3]{bp-TP-general}. Our results are independent of any concrete
such definition because the transports, we are considering here, are
independent of the particular path they are acting along (see
theorem~\ref{4-Thm3.1}).
	\end{Rem}

%	\begin{Rem}	\label{4-Rem4.3b}
%	This theorem has a version valid when the continuity of the paths
% involved is dropped. But its real power and significance is just in the
% case when all paths are assumed to be continuous.
%	\end{Rem}

	\begin{Proof}
	If $L$ is Euclidean, then, by definition~\ref{4-Defn3.4}, it admits
normal frame(s) along every $\gamma\colon J\to U$ and, consequently, according
to proposition~\ref{4-Prop4.2}, the condition~\eref{4-4.1} is valid along every
$\gamma\colon J\to U$. By theorem~\ref{4-Thm3.1}, the transport $L_{s\to
t}^{\gamma}$, $s,t\in J$ depends only on the points $x=\gamma(s)$ and
$y=\gamma(t)$ but not on the particular path $\gamma$ connecting $x,y\in U$.
Equations~\eref{4-4.3} and~\eref{4-4.4} follow from here.

	Conversely, let~\eref{4-4.1}, \eref{4-4.3}, and~\eref{4-4.4} be true for
all paths $\gamma$, $\gamma_1$, and $\gamma_2$ in $U$, the end of
$\gamma_1$ coinciding with the beginning of $\gamma_2$, and subinterval
$J'\subseteq J$.
	Meanwhile, we notice the %symbolical
 equality
	\begin{equation}	\label{4-4.5}
L^{\gamma^{-1}}=\bigl(L^\gamma\bigr)^{-1},
	\end{equation}
 $\gamma^{-1}$ being the path inverse to $\gamma$,%
\footnote{%
% Here we do not need a particular definition of $\gamma^{-1}$ (\cf
% Remark~\ref{4-Rem4.3a}). More precisely,
If
$\gamma\colon [p,q]\to U$,  and $\gamma^{-1}\colon [p',q']\to U$, for
$p,q,p',q'\in\mathbb{R}$, $p<q$, $p'<q'$, and $\gamma^{-1}(p')=\gamma(q)$,
we shall apply~\eref{4-4.5} in the form
\(
L_{p'\to q'}^{\gamma^{-1}}
   = \bigl(L_{p\to q}^{\gamma}\bigr)^{-1}
   = L_{q\to p}^{\gamma}.
\)%
}
which is a consequence of~\eref{4-4.1} and~\eref{4-4.3}.

	Let $x_0$ be arbitrarily chosen fixed point in $U$ and
$\{e_{i}^{0}\}$  an arbitrarily fixed basis in the fibre $\pi^{-1}(x_0)$
over it. In the fibre $\pi^{-1}(x)$ over $x\in U$ we define a basis
$\{e_i|_x\}$ via (cf.~\eref{4-4.2})
	\begin{equation}	\label{4-4.6}
e_i|_x := L_{s_0\to s}^{\gamma_{x_0,x}} \bigl( e_i^0 \bigr)
	\end{equation}
where $\gamma_{x_0,x}\colon J\to U$ is an arbitrary continuous path through
$x_0$ and $x$, i.e., for some $s_0,s\in J$, we have $\gamma_{x_0,x}(s_0)=x_0$
and $\gamma_{x_0,x}(s)=x$. Below we shall prove that the field $\{e_i\}$ of
bases over $U$ is normal for $L$ on $U$.

	At first, we shall prove the independence of $e_i|_x$ from the
particular continuous path $\gamma_{x_0,x}$. Let $\beta_a\colon J_a\to U$,
$a=1,2$ and $\beta_a(s_a)=x_0$ and $\beta_a(t_a)=x$ for some $s_a,t_a\in J_a$,
$a=1,2$. For definiteness, we assume $s_a\leq t_a$. (The other combinations of
ordering between $s_1$, $t_1$, $s_2,$ and $t_2$ can be considered analogously.)
Defining
 $\beta_{a}^{\prime}:=\beta_a|[s_a,t_a]$, $a=1,2$ and
using~\eref{4-4.4}, \eref{4-4.5}, \eref{4-4.3}, and~\eref{4-4.1}, we get
\[%	\begin{equation*}
L_{s_2\to t_2}^{\beta_2} \circ L_{t_1\to s_1}^{\beta_1}
  = L_{s_2\to t_2}^{\beta_2^{\prime}} \circ L_{t_1\to s_1}^{\beta_1^{\prime}}
  = L_{s_2\to t_2}^{\beta_2^{\prime}} \circ
			L_{s_1\to t_1}^{ (\beta_1^{\prime})^{-1} }
  = L_{s_0\to t_0}^{ { (\beta_1^{\prime})^{-1} }\beta_2^{\prime} }
  = \id_{\pi^{-1}(x)} ,
\]%	\end{equation*}
where
\(
{\bigl(\beta_1^{\prime}\bigr)^{-1}}\beta_2^{\prime} \colon  [s_0,t_0] \to U
\)
is the product of
 ${\bigl(\beta_1^{\prime}\bigr)^{-1}}$ and $\beta_2^{\prime}$
and we have used that, from the definition of
 ${\bigl(\beta_1^{\prime}\bigr)^{-1}}$ and $\beta_2^{\prime},$
 it is clear that
\(
\bigl( {(\beta_1^{\prime})^{-1}} \beta_2^{\prime} \bigr)(s_0)
   = \bigl( {(\beta_1^{\prime})^{-1}} \beta_2^{\prime} \bigr)(t_0)
   = x,
\)
\ie
\(
{\bigl(\beta_1^{\prime}\bigr)^{-1}}\beta_2^{\prime}
\)
is a closed path passing through $x$. Applying the last result, \eref{4-2.2},
and~\eref{4-2.3}, we obtain:
	\begin{equation*}
L_{s_2\to t_2}^{\beta_2} e_{i}^{0}
  = \Bigl( L_{s_2\to t_2}^{\beta_2} \circ L_{t_1\to s_1}^{\beta_1} \Bigr)
    \circ \Bigl( L_{s_1\to t_1}^{\beta_1}\Bigr) e_{i}^{0}
  = L_{s_1\to t_1}^{\beta_1} e_{i}^{0}.
	\end{equation*}

	Since $\beta_1$ and $\beta_2$ are arbitrary, from here we conclude
that the frame $\{e_{i}\}$, defined via~\eref{4-4.6} on $U$, is
independent from the particular path used in~\eref{4-4.6}.

	Now we shall prove that $\{e_{i}\}$ is normal for $L$ on $U$, which
will complete this proof.

	From the proof of proposition~\ref{4-Prop4.2} (compare~\eref{4-4.6}
and~\eref{4-4.2}) follows that $\{e_{i}\}$ is normal for $L$ along any path in
$U$ passing through $x_0$. Let $\gamma\colon J\to U$ be such a path, $s_0\in J$
be fixed, and $\beta\colon [0,1]\to U$ be such that $\beta(0)=x$ and
$\beta(1)=\gamma(s_0)=:x_0$. Defining $\gamma_\pm:=\gamma|J_\pm$ for
$J_\pm:=\{s\in J,\ \pm s\ge\pm s_0\}$, we conclude that $\{e_{i}\}$ is normal
for $L$ along $\beta\gamma_+$ and $\beta\gamma_{-}^{-1}$. Take, for example,
the path $\beta\gamma_{+}$. If for some
$s_{0}^{\prime},s^{\prime},s^{\ast}\in\mathbb{R}$
is fulfilled
$(\beta\gamma_{+})(s_{0}^{\prime}) = x$,
$(\beta\gamma_{+})(s^{\prime}) = \gamma(s)$, and
$(\beta\gamma_{+})(s^{\ast})   = x_0$,
then, applying~\eref{4-4.6}, \eref{4-4.3}, and~\eref{4-4.4}, we find for
$s\ge s_0$:
	\begin{multline*}
e_{i}|_{\gamma(s)}
  =  L_{s_0^{\prime}\to s^{\prime}}^{\beta\gamma_{+}} \bigl({e_i|_x}\bigr)
  =  L_{s_0^{\prime}\to s^{\prime}}^{\beta\gamma_{+}} \circ
	L_{s^{\ast}\to s_0^{\prime}}^{\beta\gamma_{+}} \bigl({e_i|_x{_0}}\bigr)
  = L_{s^{\ast}\to s^{\prime}}^{\beta\gamma_{+}} \bigl(e_i\big|_x{_0}\bigr)
\\
  =  L_{s_0\to s}^{\gamma_{+}} \circ L_{0\to 1}^{\beta}\bigl({e_i|_x{_0}}\bigr)
  =  L_{s_0\to s}^{\gamma_{+}} \bigl({e_i|_x{}}\bigr)
  =  L_{s_0\to s}^{\gamma}     \bigl({e_i|_x{}}\bigr) .
	\end{multline*}
Analogously one can prove that
\(
e_{i}|_{\gamma(s)} = L_{s_0\to s}^{\gamma} \bigl(e_i|_{x}\bigr)
\)
for $s\le s_0$ by using $\beta\gamma_{-}^{-1}$ instead of $\beta\gamma_{+}$.
So, due to~\eref{4-2.2}, the frame $\{e_i\}$ satisfies~\eref{4-3.1f} along
$\gamma$. Consequently, by corollary~\ref{4-Cor3.1*}, the frame
so\ndash constructed is normal for $L$ along $\gamma$.
	\end{Proof}

	\begin{Rem}	\label{4-Rem4.4}
	According to~\cite[proposition~3.4]{bp-LTP-general}, the
equality~\eref{4-4.3} is a consequence of~\eref{4-4.4} and the
reparametrization condition
	\begin{equation}	\label{4-4.7}
L_{s\to t}^{\gamma\circ\tau} = L_{\tau(s)\to \tau(t)}^{\gamma},
		\qquad s,t\in J''
	\end{equation}
where $J''$ is an $\mathbb{R}$\ndash interval and $\tau\colon J''\to J$ is
bijection. Hence in the formulation of theorem~\ref{4-Thm4.3} we can
(equivalently) replace the condition~\eref{4-4.3} with~\eref{4-4.7}. So, we have:
\renewcommand{\theVarThm}{\ref{4-Thm4.3}$^{\prime}$}
       \begin{VarThm}	\label{4-Thm4.3'}
	A transport $L$ is Euclidean on a path\ndash connected set
$U\subseteq B$ iff~\eref{4-4.1}, \eref{4-4.4}, and~\eref{4-4.7} are valid for every
continuous path $\gamma\colon J\to U$.
	\end{VarThm}
	\end{Rem}

	The next result is analogous to proposition~\ref{4-Prop3.10}.
According to it, a frame normal for $L$ on $U\subseteq B$, if any, can be
obtained by $L$\ndash transportation of a fixed basis over some point in $U$
to the other points of $U$.

	\begin{Prop}	\label{4-Prop4.3}
	If $L$ is a Euclidean transport on a path\ndash connected set
$U\subseteq B$ and $\{e_{i}^{0}\}$ is a given basis in $\pi^{-1}(x_0)$ for a
fixed $x_0\in U$, then the frame $\{e_{i}^{}\}$ over $U$ defined via
	\begin{equation}	\label{4-4.8}
e_{i}|_x = L_{s_0\to s}^{\gamma} \bigl( e_{i}^{0} \bigr) ,
	\end{equation}
where $\gamma\colon J\to U$ is such that $\gamma(s_0)=x_0$ and $\gamma(s)=x$
for some $s_0,s\in J$, is normal for $L$ on $U$.
	\end{Prop}

	\begin{Proof}
	By theorem~\ref{4-Thm3.1}, the basis $\{e_i|_x\}$ is independent of the
particular path $\gamma$ used in~\eref{4-4.8}. According to
theorem~\ref{4-Thm4.3}, the conditions~\eref{4-4.1}, \eref{4-4.3},
and~\eref{4-4.4} hold for $L$. Further, repeating step-by-step the last
paragraph of the proof of theorem~\ref{4-Thm4.3}, we verify that $\{e_i\}$ is
normal for $L$ on $U$.

	Alternatively, the assertion is a consequence of~\eref{4-2.20} and
proposition~\ref{4-Prop4.4} presented a few lines below.
	\end{Proof}

	A simple way to check whether a given frame is normal along
some path is provided by the following proposition.

	\begin{Prop}	\label{4-Prop4.4}
	A frame $\{e_i\}$ along $\gamma\colon J\to B$ is normal for a
linear transport $L$ in $(E,\pi,B)$, $E$ being a $C^1$ manifold, along paths
if and only if the liftings $\Hat{e}_i\colon\gamma\mapsto e_i(\cdot,\gamma)$
are constant (along $\gamma$) with respect to the derivation $D$ generated by
$L$:
	\begin{equation}	\label{4-4.9}
D^{\gamma} \Hat{e}_i = 0 .
	\end{equation}
	\end{Prop}

	\begin{Proof}
	If $\{e_i\}$ is normal for $L$ along $\gamma$, equation~\eref{4-3.1f}
is valid (see corollary~\ref{4-Cor3.1*}), so~\eref{4-4.9} follows
from~\eref{4-2.20}. If~\eref{4-4.9} holds, by virtue of~\eref{4-2.20}, its
solution is%
\footnote{%
Equation~\eref{4-4.9} is an ordinary differential equation of first order
with respect to the local components of $e_i$ (see~\eref{4-2.22}).%
}
$e_i|_{\gamma(s)} = L_{s_0\to s}^{\gamma} \bigl( e_i|_{\gamma(s_0)} \bigr)$
and consequently, by proposition~\ref{4-Prop3.10}, the frame $\{e_i\}$ is
normal along $\gamma$.
	\end{Proof}

	Recall (see the remark preceding definition~\ref{4-Defn2.2}), the
path $\gamma$ in proposition~\ref{4-Prop4.4} cannot be an arbitrary
continuous path in $B$ as it must be in the set $\pi\circ\Path^k(E)$, with
$\Path^k(E)$, $k=0,1$, being the set of $C^k$ paths in $E$. Notice, the
derivative in~\eref{4-4.9} does not require $B$ to be a manifold.

	Of course, it is true that if~\eref{4-4.9} holds in a frame $\{e_i\}$
along every path $\gamma$ in $U$, the frame $\{e_i\}$ is normal for $L$ on
$U$. But it is more natural to find a `global' version of~\eref{4-4.9}
concerning the whole set $U$, not the paths in it. Since it happens that such
a result cannot be formulated solely in terms of transports along paths, it
will be presented elsewhere.%
\index{normal frame!in vector bundles!existence theorems|)}%
\footnote{%
For this purpose is required the concept of (linear) transports along maps
(see~\cite{bp-TM-general}). Alternatively, the concept of a curvature of a
linear transport along paths can be
used~\cite{bp-LTP-Cur+Tor,bp-LTP-Cur+Tor-prop}.%
}

\section {The case of a manifold as a base}
\label{4-Sect5}

	Starting from this section, we consider some peculiarities of frames
normal for linear transports along paths in a vector bundle $(E,\pi,M)$ whose
base $M$ is a $C^1$  differentiable manifold. Besides, the bundle space $E$
will be required to be a $C^1$ manifold. This will allow links to be made
with the general results of~\cite{bp-Frames-general} concerning frames normal
for derivations of the tensor algebra of the vector space of vector fields
 over a manifold which, in particular, can be linear connections.

	The local coordinates of $x\in M$ will be denoted by $x^\mu$. Here
and below the Greek indices $\alpha,\beta,\dots,\mu,\nu,\ldots$ run
from 1 to $\dim M$ and, as usual, a summation from 1 to $\dim M$ on such
indices repeated on different levels will be assumed. The below\ndash
considered paths, like $\gamma\colon J\to M$,
are supposed to be of class $C^1$ and by $\dot\gamma(s)$ is
denoted the vector tangent to $\gamma$ at $\gamma(s)$, $s\in J$, (more
precisely at $s$), \ie $\dot\gamma$ is the vector field tangent to $\gamma$
provided $\gamma$ is injective.
By $\{E_\mu\}$ will be denoted a frame along $\gamma$ in the bundle space
tangent to $M$, \ie for every $s\in J$ the vectors
$E_1|_{\gamma(s)},\dots,E_{\dim M}|_{\gamma(s)}$ form a basis in the space
$T_{\gamma(s)}(M)$ tangent to $M$ at $\gamma(s)$. In particular, the frame
$\{E_\mu\}$ can be a coordinate one, $E_\mu|_x=\frac{\pd}{\pd
x^\mu}\big|_{x}$, in some neighborhood of $x\in\gamma(J)$. Notice, if we say
that $U$ is a neighborhood of a set $V\subseteq M$, me mean that $U$ is an
open set in $M$ containing $V$.  Otherwise by a neighborhood we understand
any open set in $M$ (which set is a neighborhood of any its point in the just
pointed sense). The transports along paths investigated below are supposed to
be of class $C^1$ on the set of $C^1$ paths in $M$.

\subsection{Normal frames for linear transports}
\label{4-Subsect5.1}

	\begin{Prop}	\label{4-Prop5.1}
Let $L$ be a linear transport along paths in $(E,\pi,M)$, $E$ and $M$ being
$C^1$ manifolds, and $L$  be Euclidean on $U\subseteq M$
(resp.\ along a $C^1$ path $\gamma\colon J\to M$). Then the matrix
$\Mat{\Gamma}$ of its coefficients has the representation
	\begin{equation}	\label{4-5.1}
\Mat{\Gamma}(s;\gamma)
   = \sum_{\mu=1}^{\dim M} \Gamma_\mu(\gamma(s)) \dot\gamma^\mu(s)
   \equiv \Gamma_\mu(\gamma(s)) \dot\gamma^\mu(s)
	\end{equation}
in any frame $\{e_i\}$ along every (resp.\ the given) $C^1$ path
$\gamma\colon J\to U$, where
\(
{\Gamma}_\mu =
  \bigl[\Sprindex[\Gamma]{j\mu}{i}\bigr]_{i,j=1}^{\dim\pi^{-1}(x)}
\)
are some matrix-valued functions, defined on an open set~$V$ containing $U$
(resp.\ $\gamma(J)$) or equal to it, and $\dot\gamma^\mu$ are the components
of $\dot\gamma$ in some frame $\{E_\mu\}$ along $\gamma$ in the bundle space
tangent to $M$, $\dot\gamma=\dot\gamma^\mu E_\mu$.
	\end{Prop}

	\begin{Proof}
	By theorem~\ref{4-Thm3.2}, the representation~\eref{4-3.3'}
is valid in $\{e_i\}$ for some matrix\ndash valued function $\Mat{F}_0$ on
$U$. Hence, if $U$ is a neighborhood, equation~\eref{4-5.1} holds for
	\begin{equation}	\label{4-5.2old}
{\Gamma}_\mu(x)
= \Mat{F}_0^{-1}(x) \bigl( E_\mu(\Mat{F}_0) |_x \bigr)
	\end{equation}
with $x\in U$.
	In the general case, \eg if $U$ is a submanifold of $M$ of dimension
less than the one of $M$, the terms $E_\mu(\Mat{F}_0)|_U$,
$\mu=1,\dots,\dim M$, in the last equality may turn to be undefined as the
matrix\ndash valued function $\Mat{F}_0$ is defined only on $U$. To overcome
this possible problem, let us take some $C^1$ matrix\ndash valued function
$\Mat{F}$, defined on an open set $V$ containing $U$ (resp.\ $\gamma(J)$)
or equal to it, such that $\Mat{F}|_U=\Mat{F}_0$. Since~\eref{4-3.3}
and~\eref{4-3.3'} depend only on the values of $\Mat{F}_0$, \ie on the ones
of $\Mat{F}$ on $U$, these equations hold also if we replace $\Mat{F}_0$ in
them with $\Mat{F}$. From the so\ndash modified equality~\eref{4-3.3'}, with
$\Mat{F}$ for $\Mat{F}_0$, we see that~\eref{4-5.1} is valid for
	\begin{equation}	\label{4-5.2}
\Gamma_\mu(x) = \Mat{F}^{-1}(x) (E_\mu(\Mat{F}))|_x
	\end{equation}
with $x\in V$.
	\end{Proof}

	Consider now the transformation properties of the matrices
$\Gamma_\mu$ in~\eref{4-5.1}.
	Let $U$ be an open set, \eg $U=M$. If we change the frame $\{E_\mu\}$
in the bundle space tangent to $M$,
$\{E_\mu\}\mapsto \{E_\mu^{\prime} = B_{\mu}^{\nu}E_\nu\}$ with
$B=\bigl[B_{\mu}^{\nu}\bigr]$ being non\ndash degenerate matrix\ndash valued
function, and simultaneously the bases in the fibres $\pi^{-1}(x)$, $x\in M$,
$\{e_i|_x\} \mapsto \{e_i^{\prime}|_x = A_{i}^{j}(x)e_j|_x\}$, then,
from~\eref{4-2.26} and~\eref{4-5.1}, we see that ${\Gamma}_\mu$ transforms
into ${\Gamma}_\mu^{\prime}$ such that
	\begin{equation}	\label{4-5.3}
{\Gamma}_\mu^{\prime}
  = B_{\mu}^{\nu} A^{-1}{\Gamma}_\nu A  + A^{-1} E'_\mu(A)
  = B_{\mu}^{\nu} A^{-1}\bigl( {\Gamma}_\nu A  + E_\nu(A) \bigr)
	\end{equation}
where $A:=\bigl[A_{i}^{j}\bigr]_{i,j=1}^{\dim\pi^{-1}(x)}$ is
non\ndash degenerate and of class $C^1$.

	\begin{Note}	\label{4-Note5.1}
	While deriving~\eref{4-5.3}, we supposed~\eref{4-5.1} to be valid
on $M$, \ie for $U=M$. If $U\not=M$, equation~\eref{4-5.1} holds only on
$U$, \ie for $\gamma\colon J\to U$. Therefore the result~\eref{4-5.3} is true
only on $U$, but in this case the frames $\{e_i\}$ and $\{e'_i\}$ \emph{must}
be defined on an open set containing or equal to $U$. This follows
from~\eref{4-2.26} in which the derivative
$\frac{\od A(s;\gamma)}{\od s} = \frac{\od A(\gamma(s))}{\od s}$
enters. To derive~\eref{4-5.3}, we have expressed  $\frac{\od A(\gamma(s))}{\od
s}$ as $(E_\mu(A))|_{\gamma(s)}\dot\gamma^\mu(s)$ which is meaningful iff $A$
is defined on a neighborhood of each point in $U$. Consequently $A$, as well as
$\{e_i\}$ and $\{e'_i\}$, must be defined on an open set
 $V\supseteq U$. For this reason, below, when derivatives like $E_\mu(A)$
appear, we admit the employed frames in the bundle space $E$ to be defined
always on some neighborhood in $M$ containing or equal to the set $U$ on which
some normal frames are investigated.
	\end{Note}

	Denoting by $\Sprindex[\Gamma]{j\mu}{i}$ the components of
${\Gamma}_\mu$, we can rewrite~\eref{4-5.3} as
	\begin{equation}	\label{4-5.4}
%	\begin{multline}	\label{4-5.4}
\Sprindex[\Gamma]{j\mu}{\prime\mspace{0.7mu} i}
  = \sum_{\nu=1}^{\dim M} \sum_{k,l=1}^{\dim\pi^{-1}(x)}
	B_{\mu}^{\nu} \bigl(A^{-1}\bigr)_{k}^{i} A_{j}^{l}
		\Sprindex[\Gamma]{l\nu}{k}
% \\
	+ \sum_{\nu=1}^{\dim M} \sum_{k=1}^{\dim\pi^{-1}(x)}
		B_{\mu}^{\nu}\bigl(A^{-1}\bigr)_{k}^{i}
		E_\nu (A_{j}^{k})  .
%	\end{multline}
	\end{equation}
Thus, we observe that the functions $\Sprindex[\Gamma]{j\mu}{i}$ are very
similar to the coefficients of a linear
connection~\cite[chapter~III, \S~7]{K&N-1}. Below, in Sect.~\ref{4-Conclusion},
we shall see that this is not accidental
(compare~\eref{4-5.1} with~\eref{4-2.30}).
These functions are also called coefficients of the transport%
\index{linear transport along paths!coefficients of}%
\index{coefficients!of linear transport along paths}
 $L$. To make a distinction between
 $\Sprindex[\Gamma]{j}{i}$ and  $\Sprindex[\Gamma]{j\mu}{i}$,
we call the former ones \emph{2-index coefficients}%
\index{linear transport along paths!coefficients!2-index}%
\index{coefficients!2-index of linear transport along paths}
of $L$ and the latter ones \emph{3\ndash index coefficients}%
\index{linear transport along paths!coefficients!3-index|defined}%
\index{coefficients!3-index of linear transport along paths|defined}
of $L$ when there is a risk of ambiguities.
	Besides, if~\eref{4-5.1} holds for every $\gamma\colon J\to U$ for a
transport $L$, then, in the general case, there are (infinitely) many
such representations unless $U$ is an open set. For instance, if~\eref{4-5.1}
is valid for some $\Gamma_\mu$, it is also true if we replace in it
$\Gamma_\mu$ with $\Gamma_\mu+G_\mu$ where the matrix\ndash valued functions
$G_\mu$ are such that $G_\mu\dot\gamma^\mu=0$ for every
$\gamma\colon J\to U$; the 3-index coefficients
$\Sprindex[\Gamma]{j\mu}{i}$ of a given linear transport $L$ admitting them
are defined \emph{uniquely} on $U\subseteq M$ by~\eref{4-5.2}
or~\eref{4-5.2old} if (and only if) $U$ is an open subset of $M$, \eg
if $U=M$.
% Hence, generally, the 3\ndash index coefficients of
% $L$ depend also on $U$ and are not unique due to, e.g., the nonuniqueness of
% $\Mat{F}$ in~\eref{4-5.2}, which is subjected only to the
% condition $\Mat{F}|_U=\Mat{F}_0$.

Note that any linear transport has 2\nobreakdash-index coefficients while
3\nobreakdash-index ones exist only for some of them; in particular
such are the Euclidean transports (see proposition~\ref{4-Prop5.1} and
theorem~\ref{4-Thm5.2} below).
% 	Through the 3-index coefficients can be defined concrete classes of,
% possibly Euclidean on some sets, linear transports along paths in a given
% vector bundle with a manifold as a base. For the purpose one should define a
% transport by the matrix~\eref{4-5.1}  of its 2\ndash index coefficients in
% which $\Gamma_\mu$ are \emph{fixed} matrix\ndash valued functions over the
% whole base $M$. In particular, if $\Gamma_\mu$ are the matrices of the
% coefficients of a linear connection in the tangent bundle over $M$, we obtain
% in this way the class of parallel transports generated by such connections in
% this bundle (see~\eref{4-2.30} and the assertion after it).

	The equation~\eref{4-5.1} is generally only a necessary, but not
sufficient condition for a frame to be normal as it is stated by the following
theorem

	\begin{Thm}	\label{4-Thm5.1}
	A $C^2$ linear transport $L$ along paths is Euclidean on a
neighborhood $U\subseteq M$ if and only if in every frame the matrix
$\Mat{\Gamma}$ of its coefficients has a representation~\eref{4-5.1} along
every $C^1$ path $\gamma$ in $U$ in which the matrix\nobreakdash-valued
functions $\Gamma_\mu$, defined on an open set containing $U$ or equal to it,
satisfy the equalities
	\begin{equation}	\label{4-5.5}
\bigl( R_{\mu\nu}(-{\Gamma}_1,\ldots,-{\Gamma}_{\dim M}) \bigr) (x)
  = 0
	\end{equation}
where $x\in U$ and
	\begin{equation}	\label{4-5.6}
R_{\mu\nu}(-{\Gamma}_1,\ldots,-{\Gamma}_{\dim M})
  := - \frac{\pd{\Gamma}_\mu}{\pd x^\nu}
     + \frac{\pd{\Gamma}_\nu}{\pd x^\mu}
     + {\Gamma}_\mu{\Gamma}_\nu
     - {\Gamma}_\nu{\Gamma}_\mu .
\end{equation}
in a coordinate frame
$\bigl\{E_\mu=\frac{\pd}{\pd x^\mu}\bigr\}$ in a neighborhood of $x$
	\end{Thm}

	\begin{Rem}	\label{4-Rem5.0}
	This result is a direct analogue
of~\cite[proposition~3.1]{bp-Frames-n+point} in the theory considered here.
	\end{Rem}

% 	\begin{Rem}	\label{4-Rem5.2}
% 	If we change the frame $\{e_i\}$,
% $\{e_i\}\mapsto\{e'_i=A_i^je_j\}$, over $U$ and
% simultaneously the frame $\{E_\mu\}$,
% $\{E_\mu\}\mapsto\{E'_\mu=B_\mu^\nu E_\nu\}$, in the tangent bundle space
% over $U$, from~\eref{4-5.6} %, \eRef[1]{7.4},
% and~\eref{4-5.3}, we get that $R_{\mu\nu}$ transform into
% 	\begin{equation}	\label{4-5.6new}
% R'_{\mu\nu} = B_\mu^\varkappa B_\nu^\lambda A^{-1} R_{\varkappa\lambda} A .
% 	\end{equation}
% Therefore the conditions~\eref{4-5.5} have an invariant character, \ie they
% are independent of the particular choice of the frames (or coordinates)
% involved in them. The last result also shows that the quantities $R_{\mu\nu}$
% depend on the frames $\{e_i\}$ and $\{E_\mu\}$ on $U$ and are completely
% independent of the their values outside $U$ (if $U\not=M$) in case they are
% defined there.% (see note~\ref[4]{Note5.1}).
% 	\end{Rem}

	\begin{Proof}
	NECESSITY.
	For a transport $L$ Euclidean on $U$ is valid~\eref{4-5.1}
due to proposition~\ref{4-Prop5.1}. Moreover, we know from the proof of this
proposition that ${\Gamma}_\mu$ admit
representation~\eref{4-5.2} for some $C^1$ non\ndash degenerate
matrix\nobreakdash-valued function $\Mat{F}$. The proof of the necessity
is completed by the following lemma.

	\begin{Lem}	\label{4-Lem5.1}
	A set of matrix-valued functions
$\{ {\Gamma}_\mu : \mu =1,\ldots,\dim M \}$, of class  $C^1$ and defined on a
neighborhood $V$, admits a representation~\eref{4-5.2} iff the
conditions~\eref{4-5.5} are fulfilled for $x\in V$.
	\end{Lem}

	\begin{Proof}[Proof of lemma~\ref{4-Lem5.1}.\@]
	A representation~\eref{4-5.2} exists iff it, considered as a
matrix linear partial differential equation of first order, has a solution
with respect to $\Mat{F}$. Rewriting~\eref{4-5.2} as
\[
\left. \frac{\pd\Mat{F}^{-1}}{\pd x^\mu} \right|_x
  = - {\Gamma}_\mu(x) \Mat{F}^{-1}(x),
\qquad x\in V,
\]
from~\cite[lemma~3.1]{bp-Frames-general} we conclude that the solutions of
this equation with respect to $\Mat{F}^{-1}$ exist iff~\eref{4-5.5} holds.
In fact, fixing some initial value $\Mat{F}^{-1}(x_0)=f_0$, we see that
	\begin{equation}	\label{4-5.7}
\Mat{F}(x)
  = f_{0}^{-1} Y^{-1}(x,x_0;-{\Gamma}_1,\ldots,-{\Gamma}_{\dim M})
	\end{equation}
where $Y(x,x_0;Z_1,\ldots,Z_{\dim M})$ is the solution of the initial-value
problem
	\begin{equation}	\label{4-5.8}
\left.\frac{\pd Y}{\pd x^\mu}\right|_x  = Z_\mu(x) Y|_x,
  \qquad
\left.Y\right|_{x=x_0} = \openone .
	\end{equation}
Here $Z_1,\ldots,Z_{\dim M}$ are continuous matrix-valued functions and
$\openone$ is the identity (unit) matrix of the corresponding size. According
to~\cite[lemma 3.1]{bp-Frames-general}, the problem~\eref{4-5.8} with
$Z_\mu=-\Gamma_\mu$ has (a unique) solution (of class $C^2$) iff the
(integrability) conditions~\eref{4-5.5} are valid.
	\end{Proof}

	SUFFICIENCY.
	Let~\eref{4-5.1} and~\eref{4-5.5} be valid. As a
consequence of lemma~\ref{4-Lem5.1}, there is a representation~\eref{4-5.2} for
${\Gamma}_\mu$ with some $\Mat{F}$. Substituting~\eref{4-5.2}
into~\eref{4-5.1}, we get
\[
\Mat{\Gamma}(s;\gamma)
  = \Mat{F}^{-1}(\gamma(s))
	\left.\frac{\pd\Mat{F}(x)}{\pd x^\mu}\right|_{x=\gamma(s)}
	\dot\gamma^\mu(s)
  = \Mat{F}^{-1}(\gamma(s))
	\frac{\od\Mat{F}(\gamma(s))}{\od s} .
\]
	So, by theorem~\ref{4-Thm3.2} (see~\eref{4-3.3'} for
$\Mat{F}_0=\Mat{F}|_U$), the considered transport $L$ along paths is
Euclidean.
	\end{Proof}

	The just-proved theorem~\ref{4-Thm5.1} expresses a very important
practical necessary and sufficient condition for existence of frames normal
on neighborhoods because the conditions~\eref{4-5.1} and~\eref{4-5.5} are
easy to check for a given linear transport along paths in bundles with a
differentiable manifold as a base.

	Now, combining~\eref{4-3.1c} and~\eref{4-5.1}, applying
corollary~\ref{4-Cor3.1*}, and using the arbitrariness of
$\gamma$, we can formulate the following essential result.

	\begin{Prop}	\label{4-Prop5.2}
	A necessary and sufficient condition for a frame to be normal on a
neighborhood $U\subseteq M$ for a Euclidean  linear transport on U along
paths in $(E,\pi,M)$ is the vanishment of its 3\ndash index coefficients, \ie
	\begin{equation}	\label{4-5.9}
{\Gamma}_\mu(x)
  := \bigl[\Sprindex[\Gamma]{j\mu}{i}\bigr]_{i,j=1}^{\dim\pi^{-1}(x)}
  = 0
	\end{equation}
for every $x\in U$, where ${\Gamma}_\mu(x)$ define
the (2\ndash index) coefficients of the transport via~\eref{4-5.1}.
	\end{Prop}

	Now we are going to find an analogue of theorem~\ref{4-Thm5.1} when
the neighborhood $U\subseteq M$ in it is replaced with a submanifold of the
base $M$.

	Let $N$ be a submanifold of $M$ and $L$ a linear transport along paths
in $(E,\pi,M)$ which is Euclidean on $N$. Let the $C^1$ matrix\ndash valued
function $\Mat{F}_0$ determines the coefficients' matrix of $L$
via~\eref{4-3.3'}. Suppose $p_0\in N$ and  $(V,x)$ is a chart of $M$ such that
$V\ni p_0$ and the local coordinates of every $p\in N\cap V$ are
 $x(p)=( x^1(p),\dots,x^{\dim N}(p),t_0^{\dim N+1},\dots,t_0^{\dim M} )$,
where $t_0^\rho$, $\rho=\dim N+1,\dots,\dim M$, are constant numbers.%
\footnote{%
We are using the definition of a submanifold presented
in~\cite[p.~227]{Bruhat}.%
}

	In the chart $(V,x)$, we have
\(
\frac{\od \Mat{F}_0(\gamma(s))}{\od s}
= \sum_{\alpha=1}^{\dim N}
	\frac{\pd \Mat{F}_0}{\pd x^\alpha}\Big|_{\gamma(s)}
		\dot\gamma^\alpha(s)
\),
with $\gamma^\mu:=x^\mu\circ\gamma$, for every  $C^1$ path $\gamma\colon
J\to N$ and $s\in J$. From here and~\eref{4-3.3'}, it follows
that~\eref{4-5.1} holds for
	\begin{equation}	\label{4-5.11}
\Gamma_\alpha(p)
= \Mat{F}_0^{-1}(p) \frac{\pd \Mat{F}_0}{\pd x^\alpha} \Big|_p,
\qquad \alpha=1,\dots,\dim N
	\end{equation}
and \emph{arbitrary} $\Gamma_{\dim N+1},\dots,\Gamma_{\dim M}$ since in the
coordinates $\{x^\mu\}$ is fulfilled $\gamma^\rho(s)=t_0^\rho=\const$ and
hence
	\begin{equation}	\label{4-5.12}
\dot\gamma^{\dim N+1}=\cdots=\dot\gamma^{\dim M} \equiv 0.
	\end{equation}
Comparing~\eref{4-5.11} with~\eref{4-5.2old} for
$E_\mu=\frac{\pd}{\pd x^\mu}$, we conclude that $\Gamma_\alpha$, given
via~\eref{4-5.11}, are exactly the first $\dim N$ of the matrices
$\Gamma_\mu=[\Sprindex[\Gamma]{j\mu}{i}]$ of the 3\ndash index
coefficients of the transport $L$ in the pair of frames
$\bigl( \{e_i\},\bigl\{\frac{\pd}{\pd x^\mu}\bigr\} \bigr)$.
As we said, the rest of the 3\ndash index coefficients of $L$ (on $N$) are
completely arbitrary. In particular, one can choose them according
to~\eref{4-5.2},
	\begin{equation}	\label{4-5.13}
\Gamma_\rho(p) = \Mat{F}^{-1}(p) \frac{\pd\Mat{F}}{\pd x^\rho},
\qquad \rho=\dim N+1,\dots,\dim M,
\quad  \Mat{F}|_N=\Mat{F}_0,
	\end{equation}
which leads to the validity of~\eref{4-5.2} in every frame, or, if the
representation~\eref{4-5.1} holds for every $\gamma\colon J\to M$ (this does
not mean that $L$ is Euclidean on $M$!), the matrices $\Gamma_\rho$ can be
identified with the ones appearing in~\eref{4-5.1} in the frame
$\bigl\{\frac{\pd}{\pd x^\mu}\bigr\}$.

	If $\{x^{\prime\,\mu}\}$  are other coordinates on $V$ like
$\{x^\mu\}$, \ie $x^{\prime\,\rho}(p)=\const$ for $p\in N\cap V$ and
$\rho=\dim N+1,\dots,\dim M$, the change
$\{x^\mu\} \mapsto \{x^{\prime\,\mu}\}$, combined with
$\{e_i\}\mapsto\{e'_i=A_i^j e_j\}$ leads to
	\begin{equation}	\label{4-5.13new}
\Gamma_\alpha\mapsto \Gamma'_\alpha
= B_\alpha^\beta A^{-1}\Gamma_\beta A
	+ A^{-1}\frac{\pd A}{\pd x^{\prime\,\alpha}},
\quad
B_\alpha^\beta  :=  \frac{\pd x^\beta}{\pd x^{\prime\,\alpha}},
\quad
\alpha,\beta=1,\dots, \dim N
	\end{equation}
on $N\cap V$. So, equation~\eref{4-5.3} remains valid only for frames
$\{E_\mu\}$ normal on $N$. But using the arbitrariness of $\Gamma_\rho$,
we can force~\eref{4-5.3} to hold on $N$ for arbitrary frames defined on a
neighborhood of $N$.

	The above discussion implies that the condition~\eref{4-5.5} in
theorem~\ref{4-Thm5.1}, when applied on a submanifold $N$, imposes
restrictions on the transport $L$ as well as ones on the `inessential'
3\ndash index coefficients of $L$, like $\Gamma_\rho$ above, or on the
matrix\ndash valued function $\Mat{F}$ entering in~\eref{4-5.2} or
in~\eref{4-5.13}. Since the restrictions of the last type are not connected
with the transport $L$, below we shall `repair' theorem~\ref{4-Thm5.1} on
submanifolds in such a way as   to exclude them from the final results.

	\begin{Thm}	\label{4-Thm5.2}
A linear transport $L$ along paths is Euclidean on a submanifold $N$ of $M$
if and only if in every frame $\{e_i\}$, in the bundle space over $N$, the
matrix of its coefficients has a representation~\eref{4-5.1} along every
$C^1$ path in $N$ and, for every $p_0\in N$ and a chart $(V,x)$ of  $M$
such that $V\ni p_0$ and
$x(p)=( x^1(p),\dots,x^{\dim N}(p),t_0^{\dim N+1},\dots,t_0^{\dim M} )$
for every $p\in N\cap V$ and constant numbers
$t_0^{\dim N+1},\dots,t_0^{\dim M}$, the equalities
	\begin{equation}	\label{4-5.14}
\bigl(  R^N_{\alpha\beta}(-\Gamma_1,\ldots,-\Gamma_{\dim N}) \bigr) (p) = 0,
\qquad
\alpha,\beta=1,\dots,\dim N
	\end{equation}
hold for all $p\in N\cap V$ and
%	\begin{multline}	\label{4-5.15}
	\begin{equation}	\label{4-5.15}
R^N_{\alpha\beta}(-\Gamma_1,\ldots,-\Gamma_{\dim N})
%\\
:= R_{\alpha\beta}(-\Gamma_1,\ldots,-\Gamma_{\dim M})
 =   - \frac{\pd \Gamma_\alpha} {\pd x^\beta}
     - \frac{\pd \Gamma_\beta}  {\pd x^\alpha}
     + \Gamma_\alpha \Gamma_\beta
     - \Gamma_\beta  \Gamma_\alpha .
	\end{equation}
%	\end{multline}
Here $\Gamma_1,\ldots,\Gamma_{\dim N}$ are first $\dim N$ of the matrices of
the 3-index coefficients of $L$ in the coordinate frame
$\bigl\{\frac{\pd}{\pd x^\mu}\bigr\}$ in the tangent bundle space over
$N\cap V$. They are uniquely defined via~\eref{4-5.11}.
	\end{Thm}

	\begin{Rem}	\label{4-Rem5.1}
	In the theory considered here, this result is a direct analogue
of~\cite[theorem~3.1]{bp-Frames-general}.
	\end{Rem}

% 	\begin{Rem}	\label{4-Rem5.1new}
% 	It is intuitively clear, generally not all of the
% equations~\eref{4-5.14} are independent. One can expect only
% $(\dim N)[(\dim N)-1]/2$ of them to be independent because of
% $R_{\mu\nu}=-R_{\nu\mu}$, due to~\eref{4-5.6}.
% 	\end{Rem}

	\begin{Rem}	\label{4-Rem5.5}
	This theorem is, in fact, a special case of theorem~\ref{4-Thm5.1}:
if in the latter theorem we put $U=N$, restrict the transport $L$ to the
bundle $(\pi^{-1}(N),\pi|_{\pi^{-1}(N)},N)$, replace $M$ with $N$, and notice
that $\{x^1,\dots,x^{\dim N}\}$ provide an internal coordinate system on $N$,
we get the former one.
% with $E_\alpha=\pd/\pd x^\alpha$.
 Because of the importance of the result obtained, we call it `theorem' and
present below its independent proof.
	\end{Rem}

	\begin{Proof}
	If $L$ is Euclidean on $N$, equation~\eref{4-5.1} holds in every
frame on $N$ (proposition~\ref{4-Prop5.1}); in particular it is valid in the
frame $\bigl\{\frac{\pd}{\pd x^\mu}\bigr\}$, induced by the chart $(V,x)$, in
which, as was proved above, equation~\eref{4-5.11} is satisfied. The
substitution of~\eref{4-5.11} into~\eref{4-5.15} results
in~\eref{4-5.14}. Conversely, let~\eref{4-5.1} for $\gamma\colon J\to N$
and~\eref{4-5.14} be valid. By lemma~\ref{4-Lem5.1} with $N$ for $M$,
from~\eref{4-5.14} follows the existence of a representation~\eref{4-5.11}
for some matrix\ndash valued function $\Mat{F}_0$ on $N$.
Substituting~\eref{4-5.11} into~\eref{4-5.1} and using that $\gamma$ is a
path in $N$ and~\eref{4-5.12} is valid, in the frame
$\bigl\{\frac{\pd}{\pd x^\mu}\bigr\}$, we obtain:
	\begin{multline*}
\Mat{\Gamma}(s;\gamma)
= \Gamma_\mu(\gamma(s)) \dot\gamma^\mu(s)
= \sum_{\alpha=1}^{\dim N} \Gamma_\alpha(\gamma(s)) \dot\gamma^\alpha(s)
= \Mat{F}^{-1}_0(\gamma(s))
	\sum_{\alpha=1}^{\dim N}
	\frac{\pd\Mat{F}_0}{\pd x^\alpha} \bigg|_{\gamma(s)}
	\dot\gamma^\alpha(s)
\\
= \Mat{F}^{-1}_0(\gamma(s))
	\frac{\pd\Mat{F}}{\pd x^\mu} \bigg|_{\gamma(s)}
	\dot\gamma^\mu(s)
= \Mat{F}^{-1}_0(\gamma(s))
	\frac{\od\Mat{F}(\gamma(s))}{\od s}
= \Mat{F}^{-1}_0(\gamma(s))
	\frac{\od\Mat{F}_0(\gamma(s))}{\od s}
	\end{multline*}
where $\Mat{F}$ is a $C^1$ matrix-valued function defined on an open set
containing $N$ or equal to it and such that $\Mat{F}|_N=\Mat{F}_0$. Thus, by
theorem~\ref{4-Thm3.2}, the transport $L$ is Euclidean on $N$.
	\end{Proof}

	\begin{Cor}	\label{4-Cor5.1}
	Every linear transport along paths in a vector bundle whose base and
bundle spaces are $C^1$ manifolds, is Euclidean at every single point or
along every path without self\ndash intersections.
	\end{Cor}

	\begin{Proof}
See theorem~\ref{4-Thm5.2} for $\dim N=0,1$, in which cases
$R^N_{\alpha\beta}\equiv0$.
	\end{Proof}

	It should be noted, the last result agrees completely with
proposition~\ref{4-Prop4.1} and corollary~\ref{4-Cor4.1}.

% \section
\subsection
% [Normal frames for derivations in vector bundles with\\ a manifold as a base]
% {Normal frames for derivations in vector bundles with   a manifold as a base}
{Normal frames for derivations}
\label{4-Sect6}

	For a general bundle $(E,\pi,B)$ whose bundle space $E$ is $C^1$
manifold, we call a frame $\{e_i\}$ \emph{normal on}%
\index{normal frame!for derivation along paths|(}%
\index{derivation along paths!normal frame for|(}
	$U\subseteq B$ (resp.\ along $\gamma\colon J\to M$) for a
derivation $D$ along paths (resp.\ $D^\gamma$ along $\gamma$) (see
definition~\ref{4-Defn2.2}) if $\{e_i\}$ is normal on $U$ (resp.\ along
$\gamma$) for the linear transport $L$ along paths generating it
by~\eref{4-2.18} (see proposition~\ref{4-Prop2.6}). We can also  equivalently
define a frame normal for $D$ (resp.\ $D^\gamma$) as  one in which the
components of $D$ (resp.\ $D^\gamma$)  vanish (see the proof of
proposition~\ref{4-Prop2.6},  and
corollary~\ref{4-Cor3.1*}). A derivation admitting normal frame(s) is called
\emph{Euclidean}.%
\index{derivation along paths!Euclidean}

	In connection with concrete physical applications, far more
interesting case is the case of a bundle $(E,\pi,M)$ with a differentiable
manifold $M$ as a base. The cause for this is the existence of natural
structures over $M$, \eg the different tensor bundles and the tensor algebra
over it. Below we concentrate on this particular case.

	\begin{Defn}	\label{4-Defn6.1}
\index{derivation along tangent vector fields|defined[\ff{}]}
	A \emph{derivation} over an open set $V\subseteq M$ or in
$(E,\pi,M)|_V$ \emph{along tangent vector fields} is a map $\mathcal{D}$
assigning to every tangent vector field $X$ over $V$ a linear map
	\begin{equation}	\label{4-6.1}
\mathcal{D}_X
  \colon
\Sec^1\bigl( (E,\pi,M)|_V \bigr) \to \Sec^0\bigl( (E,\pi,M)|_V \bigr) ,
	\end{equation}
called a \emph{derivation along} $X$, such that
	\begin{equation}	\label{4-6.2}
\mathcal{D}_X (f\cdot\sigma)
  = X(f)\cdot\sigma + f\cdot\mathcal{D}_X(\sigma)
	\end{equation}
for every $C^1$ section $\sigma$ over $V$ and every $C^1$ function
$f\colon V\to\mathbb{C}$.
	\end{Defn}

	Obviously (see definition~\ref{4-Defn2.2}), if $\gamma\colon J\to V$
is a $C^1$ path, the map
$ \overline{D}\colon\hat{\sigma}\mapsto \overline{D}\hat{\sigma} $, with
\(
\overline{D}\hat{\sigma} \colon \gamma\mapsto \overline{D}^\gamma \hat{\sigma},
\)
where
\(
\overline{D}^\gamma\hat{\sigma}\colon s
	\mapsto \overline{D}_{s}^{\gamma}\hat{\sigma}
\)
is defined via
	\begin{equation}	\label{4-6.3}
\overline{D}_s^\gamma(\hat{\sigma})
 = \bigl( (\mathcal{D}_X\sigma)|_{X=\dot\gamma} \bigr) (\gamma(s)),
\qquad
\hat{\sigma}\colon\gamma\mapsto \sigma\circ\gamma,
	\end{equation}
is a derivation along paths on the set of $C^1$ liftings generated by
sections of $(E,\pi,M)|_V$. From Sect~\ref{4-Sect2}, we know that along
paths without self\ndash intersections every derivation along paths generates
a derivation of the sections of $(E,\pi,M)$ (see~\eref{4-2.17}
and~\eref{4-2.17*}). Thus to any derivation $\mathcal{D}$ along (tangent)
vector fields on $V$ there corresponds, via~\eref{4-6.3}, a natural
derivation $D$ along the paths in $V$ on the set of liftings generated by
sections.  These facts are a hint for the possibility to introduce `normal'
frames for $\mathcal{D}$. This can be done as follows.

	Let $\{e_i\}$ be a  $C^1$ frame in $\pi^{-1}(V)$.
We define the \emph{components}%
\index{derivation along tangent vector fields!components of|defined}%
\index{components!of derivation along tangent vector fields|defined}
	or \emph{ (2\ndash index) coefficients}%
\index{derivation along tangent vector fields!2-index coefficients of|defined}%
\index{coefficients!2-index of derivation along tangent vector fields|defined}
$\Sprindex[\text{$\Gamma_X$}]{j}{i} \colon  V\to\mathbb{C}$
of $\mathcal{D}_X$ by the expansion (cf.~\eref{4-2.29})
	\begin{equation}	\label{4-6.4}
\mathcal{D}_X e_i = \Sprindex[\text{$\Gamma_X$}]{i}{j}\, e_j .
	\end{equation}
So
$\Gamma_X:=\bigl[ \Sprindex[\text{$\Gamma_X$}]{i}{j} \bigr]$
is the \emph{matrix} %
\index{derivation along tangent vector fields!matrix of|defined}%
\index{matrix!of derivation along tangent vector fields|defined}%
of $\mathcal{D}_X$ in $\{e_i\}$.

	Applying~\eref{4-6.2} to $\sigma=\sigma^ie_i$ and using the linearity
of $\mathcal{D}_X$, we get the explicit expression (cf.~\eref{4-2.22})
	\begin{equation}	\label{4-6.5-0}
\mathcal{D}_X(\sigma)
  = \bigl( X(\sigma^i)
	+ \Sprindex[\text{$\Gamma_X$}]{j}{i}\, \sigma^j \bigr) e_i .
	\end{equation}

	A simple verification proves that the change
 $\{e_i\}\mapsto\{e_{i}^{\prime}=A_{i}^{j}e_j\}$, with a non\ndash degenerate
$C^1$ matrix\nobreakdash-valued function $A=\bigl[A_{i}^{j}\bigr]$, leads to
(cf.~\eref{4-2.26})
	\begin{equation}	\label{4-6.5}
\Gamma_X
  := \bigl[ \Sprindex[\text{$\Gamma_X$}]{j}{i} \bigr]
  \mapsto
\Gamma_{X}^{\prime}
  := \bigl[ \Sprindex[\text{$\Gamma_X^\prime$}]{j}{i} \bigr]
  = A^{-1}\Gamma_XA +A^{-1}X(A),
	\end{equation}
where $X(A):=\bigl[X(A_{i}^{j})\bigr]$. Conversely, if a geometrical object
with components $\Sprindex[\text{$\Gamma_X$}]{j}{i}$ is given in a frame
$\{e_i\}$ and a change $\{e_i\}\mapsto\{e_{i}^{\prime}=A_{i}^{j}e_j\}$ implies
the transformation~\eref{4-6.5}, then there exists a unique derivation along
$X$, defined via~\eref{4-6.5-0}, whose components in $\{e_i\}$ are exactly
$\Sprindex[\text{$\Gamma_X$}]{j}{i}$ (\cf proposition~\ref{4-Prop2.5new}).

	Below, for the sake of simplicity, we take $V=M$, \ie the derivations
are over the whole base $M$.

	\begin{Defn}	\label{4-Defn6.2}
\index{normal frame!for derivation along tangent vector fields|defined[\ff{}]}%
\index{derivation along tangent vector fields!normal frame for|defined[\ff{}]}
	A \emph{frame} $\{e_i\}$, defined on an open set containing $U$ or
equal to it, is called
\emph{normal for a derivation $\mathcal{D}$ along tangent vector fields
(resp.\ for $\mathcal{D}_X$ along a given tangent vector field $X$) on}
$U$ if in $\{e_i\}$ the components of $\mathcal{D}$ (resp.\ $\mathcal{D}_X$)
vanish on $U$ for every (resp.\ the given) tangent vector field $X$.
	\end{Defn}

	If $\mathcal{D}$ (resp.\ $\mathcal{D}_X$) admits frames normal on
$U\subseteq M$, we call it \emph{Euclidean}%
\index{derivation along tangent vector fields!Euclidean|defined[\ff{}]}
 on $U$. A number of results, analogous to those of
sections~\ref{4-Sect3}--\ref{4-Subsect5.1}, can be proved for such derivations.
Here we shall mention only a few of them.

	\begin{Prop}[\normalfont
			cf.\ theorem~\protect{\ref{4-Thm3.2}}\bfseries]
	\label{4-Prop6.1}
	A derivation $\mathcal{D}$ along vector fields admits frame(s) normal
on $U\subseteq M$ iff in every frame its matrix on $U$ has the form
	\begin{equation}	\label{4-6.6}
\Gamma_X|_U = \bigl( F^{-1}  X(F) \bigr)_U
	\end{equation}
where $F$ is a  $C^1$ non\ndash degenerate matrix-valued function defined on
an open set containing $U$.
	\end{Prop}

	\begin{Proof}
	If $\{e_i^\prime\}$ is normal on $U$  for $\mathcal{D}$,
then~\eref{4-6.6} with $F=A^{-1}$ follows from~\eref{4-6.5} with
$\Gamma_X^\prime|_U=0$. Conversely, if~\eref{4-6.6} holds,
then~\eref{4-6.5} with $A=F^{-1}$ yields $\Gamma_{X}^{\prime}|_U=0$.
	\end{Proof}

	\begin{Prop}[\normalfont
			cf.\ corollary~\protect{\ref{4-Cor3.3}}\bfseries]
	\label{4-Prop6.2}
	The frames normal on a set $U\subseteq M$ for a Euclidean derivation
along vector fields (resp.\ given vector field $X$) are connected by linear
transformations whose matrices $A$ are constant (resp.\ $X(A)=0$) on
$U$.

	\end{Prop}
	\begin{Proof}
	The result is a consequence of~\eref{4-6.5} for
$\Gamma_X=\Gamma_{X}^{\prime}=0$.
	\end{Proof}

	\begin{Defn}	\label{4-Defn6.3}
\index{derivation along tangent vector fields!linear|defined[\ff{}]}
	A \emph{derivation $\mathcal{D}$  along (tangent) vector fields} is
called \emph{linear on} $U$ if in one (and hence in any) frame its components
admit the representation
	\begin{equation}	\label{4-6.7}
\Sprindex[\text{$\Gamma_X$}]{j}{i}(x)
  = \Sprindex[\Gamma]{j\mu}{i}(x) X^\mu(x)
\qquad\text{or}\quad
\Gamma_X=\Gamma_\mu X^\mu
	\end{equation}
where $x\in U$,
\(
\Gamma_\mu
  = \bigl[ \Sprindex[\Gamma]{j\mu}{i}(x) \bigr]_{i,j=1}^{\dim\pi^{-1}(x)}
\)
are matrix-valued functions on $U$, and $X^\mu$ are the local components of
a vector field $X$ in some frame $\{E_\mu\}$ of tangent vector fields,
$X=X^\mu E_\mu$.
	\end{Defn}

	\begin{Rem}	\label{4-Rem6.1}
\index{derivation along tangent vector fields!linear|defined[\ff{}]}
	The invariant definition of a derivation linear on $U$ is via the
equation
	 \begin{equation}	\label{4-6.8}
\mathcal{D}_{fX+gY} = f\mathcal{D}_X + g\mathcal{D}_Y
	\end{equation}
where $f,g\colon U\to\mathbb{C}$ and  $X$ and $Y$ are tangent vector fields over
$U$. But for the purposes of this work the above definition is more
suitable. Comparing definitions~\ref{4-Defn6.1} and~\ref{4-Defn6.3} (see
also~\eref{4-6.8}) with~\cite[p.~74, definition~2.51]{Poor}, we see that a
derivation along tangent vector fields is linear iff it is a covariant
derivative operator%
\index{covariant derivative operator}
in $(E,\pi,B)$. Therefore the concepts linear derivation
along tangent vector fields and covariant derivative operator coincide.
	\end{Rem}

	We call $\Sprindex[\Gamma]{j\mu}{i}$ \emph{3-index coefficients}%
\index{derivation along tangent vector fields!3-index coefficients of|defined}%
\index{coefficients!3-index of derivation along tangent vector fields|defined}%
\index{derivation along tangent vector fields!coefficients of|defined}%
\index{coefficients!of derivation along tangent vector fields|defined}
 of
$\mathcal{D}$ or simply coefficients if there is no risk of misunderstanding.
It is trivial to check that under changes of the frames they transform
according to~\eref{4-5.4}. It is easy to verify that to every linear
derivation $\mathcal{D}$ there corresponds a unique derivation along paths
or linear transport along paths whose 2\nobreakdash-index coefficients are
given via~\eref{4-5.1} with
$\Gamma_\mu:=\bigl[\Sprindex[\Gamma]{j\mu}{i}\bigr]$ being the matrices of
the 3\ndash index coefficients of $\mathcal{D}$.%
\footnote{%
One can verify that the action of the derivation along paths induced
by $\mathcal{D}$ on the liftings generated by sections is given
by~\eref{4-6.3}.%
}
Conversely, to any such transport or
derivation along paths there corresponds a unique linear derivation along
tangent vector fields with components ((2\nobreakdash-index) coefficients)
given by~\eref{4-6.7}, \ie with the same 3\nobreakdash-index coefficients.
So,
\emph{there is a bijective correspondence between the sets of linear
derivations along tangent vector fields and derivations (or linear
transports) along paths whose (2\nobreakdash-index) coefficients admit the
representation~\eref{4-5.1}}.
	It should be emphasized, if the above discussion is restricted to a
subset $U$, \ie only for paths lying entirely in $U$, it remains valid iff
$U$ is an open set in $M$.
% Otherwise, if $U$ is not an open set, the
% correspondence between derivations along tangent vector fields and
% derivations or linear transports along paths via their 3\ndash index
% coefficients is surjective in the right direction.%
% \footnote{%
% If $\Gamma_\mu$ and $\overline{\Gamma}_\mu$ are the matrices of the 3-index
% coefficients of $\mathcal{D}$ and $\overline{\mathcal{D}}$, then they define
% a single derivation along paths in $U$ iff
% $(\overline{\Gamma}_\mu - \Gamma_\mu)\dot\gamma^\mu=0$ for all $C^1$ paths
% $\gamma\colon J\to U$.%
% }
% In the opposite direction it is injective, if the 3\ndash index
% coefficients of the derivations along paths are fixed, or/and to a single
% derivation along paths may correspond different derivations along tangent
% vector fields, if the arbitrariness of the 3\ndash index coefficients of the
% former derivations is taken into account. This remark is important when the
% normal frames in the both cases are compared.

	\begin{Prop}	\label{4-Prop6.3}
	A derivation along tangent vector fields is Euclidean on $U$ iff it
is linear on $U$ and, in every frame $\{e_i\}$ over $U$ in the bundle space
and every local coordinate frame $\bigl\{E_\mu=\frac{\pd}{\pd x^\mu}\bigr\}$
over $U$ in the tangent bundle space over $U$, the matrices
$\Sprindex[\Gamma]{\mu}{}$ of its 2\nobreakdash-index coefficients have the
form~\eref{4-5.2} for some non\ndash degenerate $C^1$ matrix\ndash valued
function $\Mat{F}$ on $U$.
	\end{Prop}

	\begin{Proof}
	The result is a corollary from proposition~\ref{4-6.1} as
$X=X^\mu\frac{\pd}{\pd x^\mu}$ and~\eref{4-6.6} imply~\eref{4-6.7} with
$\Gamma_\mu:=[\Sprindex[\Gamma]{j\mu}{i}]=F^{-1}\frac{\pd F}{\pd x^\mu}$.
	\end{Proof}

	\begin{Thm}[\normalfont
			cf.\ theorem~\protect{\ref{4-Thm5.1}}\bfseries]
	\label{4-Thm6.1}
	Frames normal on a neighborhood $U$ for a derivation $\mathcal{D}$
along vector fields exist iff it is linear on $U$ and its
3\nobreakdash-index coefficients satisfy the conditions~\eref{4-5.5} on $U$.
	\end{Thm}

	\begin{Proof}
	By proposition~\ref{4-Prop6.3},  a derivation $\mathcal{D}$ along
vector fields is Euclidean iff~\eref{4-5.2} holds for some $\Mat{F}$ which,
according to lemma~\ref{4-Lem5.1}, is equivalent to~\eref{4-5.5}.
	\end{Proof}

	\begin{Prop}[\normalfont
			cf.\ proposition~\protect{\ref{4-Prop5.2}}\bfseries]
	\label{4-Prop6.4}
	A frame is normal on a set $U$ for some linear derivation along
tangent vector fields iff the derivation's 3\nobreakdash-index coefficients
vanish on $U$.
	\end{Prop}

	\begin{Proof}
	This result is a corollary of definition~\ref{4-Defn6.2},
equation~\eref{4-6.7} and the arbitrariness of $X$ in it.
	\end{Proof}

% 	\begin{Rem}	\label{4-Rem6.2}
% 	The arbitrariness of $U$ in this proposition does not contradict
% remark~\ref{4-Rem5.4} as the restriction of a derivation along tangent vector
% fields to $U$ is along vector fields on $U$, \ie $X_x\in T_x(M)$ with
% $x\in U$.
% 	\end{Rem}

	In this way we have proved the
\emph{existence of a bijective mapping between the sets of
Euclidean derivations along paths
and Euclidean linear transports along paths}.
It is given via the (local) coincidence of their 3\nobreakdash-index
coefficients in some (local) frame.  Moreover, the normal frames for the
corresponding objects of these sets coincide. %
\index{normal frame!for derivation along paths|)}%
\index{derivation along paths!normal frame for|)}
	What concerns  the frames normal for Euclidean derivations along
tangent vector fields, in them, by proposition~\ref{4-Prop6.4}, vanish not
only their 2\ndash index coefficients, but also the 3\ndash index ones. Hence
the set of these frames is, generally, a subset of the one of frames normal
for derivations or linear transports along paths.
% More details on frames in
% which the 3\ndash index coefficients of a derivation or transport vanish
% will be presented in Sect.~\ref{4-Sect12}.

\section{Strong normal frames}
	\label{4-Sect12}

	Let $M$ be a manifold and $(T(M),\pi,M)$ the tangent bundle over it.
Let $\nabla$ and $\mathsf{P}$ be, respectively, a linear connection on $M$
and the parallel transport along paths in $(T(M),\pi,M)$ generated by
$\nabla$ (see~\eref{4-2.30} and the statement after it). Suppose $\nabla$
and $\mathsf{P}$ admit frames normal on a set $U\subseteq M$. Here a natural
question arises:  what are the links between both types of normal frames, the
ones normal for $\nabla$ on $U$ and the ones for $\mathsf{P}$ on $U$?

	Recall,
% (see definition~\Ref[1]{Defn3.3} and~\ref{4-Defn3.1'} and
% equation~\eReftag[4]{10.5'}),
if $\Sprindex[\Gamma]{jk}{i}$ are the
coefficients of $\nabla$ in a frame $\{E_i\}$, the frame $\{E_i\}$ is normal on
$U\subseteq M$ for  $\nabla$ or $\mathsf{P}$ iff respectively
	\begin{gather}	\label{4-12.1}
\Sprindex[\Gamma]{jk}{i}(p) = 0
		\\	\label{4-12.2}
\Sprindex[\Gamma]{j}{i} (s;\gamma)
	= \Sprindex[\Gamma]{jk}{i}(\gamma(s)) \dot\gamma^k(s)
= 0
	\end{gather}
for every $p\in U$, $\gamma\colon J\to U$, and $s\in J$. Two simple but quite
important conclusions can be made from these equalities:
	(i) The
frames normal for $\nabla$ are normal for $\mathsf{P}$, the converse being
generally not valid, and
	(ii) in a frame normal for $\nabla$
vanish the 2\ndash index as well as the 3\ndash index coefficients of
$\mathsf{P}$.

	\begin{Defn}	\label{4-Defn12.1}
	Let $\mathsf{P}$ be a parallel transport in $(T(M),\pi,M)$ and
$U\subseteq M$. A \emph{frame} $\{E_i\}$, defined on an open set containing
$U$, is called \emph{strong normal on $U$ for} $\mathsf{P}$ if
the 3\ndash index coefficients of $\mathsf{P}$ in $\{E_i\}$ vanish on $U$.
Respectively, $\{E_i\}$ is \emph{strong normal along} $g\colon Q\to M$ if it
is strong normal on $g(Q)$.
	\end{Defn}

	Obviously, the set of frames strong normal on $U$ for a parallel
transport $\mathsf{P}$ coincides with the set of frames normal for the linear
connection $\nabla$ generating $\mathsf{P}$.

	The above considerations can be generalized  directly to linear
transports for which 3\ndash index coefficients exist and are fixed.

	\begin{Defn}	\label{4-Defn12.2}
	Let $E$ and $M$ be  $C^1$ manifolds, $U\subseteq M$, and $(E,\pi,M)$
be a vector bundle over $M$. Let $L$  (resp.\ $D$) be a linear transport
(resp.\ derivation) along paths in $(E,\pi,M)$ admitting 3\ndash index
coefficients on $U$ which are supposed to be fixed, \ie its coefficient
matrix is of the form
	\begin{equation}	\label{4-12.3}
\Mat{\Gamma}(s;\gamma) = \Gamma_\mu(\gamma(s)) \dot\gamma^\mu(s)
	\end{equation}
in every pair of frames $\{e_i\}$ in $E$ and $\{E_\mu\}$ in $T(M)$ defined on
an open set containing $U$ or equal to it, where $\gamma\colon J\to U$ is of
class $C^1$ and $\Gamma_\mu:=[\Sprindex[\Gamma]{j\mu}{i}]$ are the (fixed)
matrices of the 3\ndash index coefficients of $L$.
	A \emph{frame} $\{e_i\}$, defined on an open set containing $U$ or
equal to it, is called \emph{strong normal on $U$ for} $L$ (resp.\ $D$), if
in the pair $(\{e_i\},\{E_\mu\})$ for some (and hence any) $\{E_\mu\}$ the
3\ndash index coefficients of $L$ vanish on $U$.
	Respectively, $\{e_i\}$ is \emph{strong normal along} $g\colon Q\to
M$ if it is strong normal on $g(Q)$.
	\end{Defn}

	So, a frame $\{e_i\}$ is strong normal or normal on  $U$ if
(cf.~\eref{4-12.1} and~\eref{4-12.2}) respectively
	\begin{gather}	\label{4-12.4}
\Gamma_\mu(x) = 0
		\\	\label{4-12.5}
\Mat{\Gamma}(s;\gamma) = \Gamma_\mu(\gamma(s)) \dot\gamma^\mu(s) = 0
	\end{gather}
for every $x\in U$, $\gamma\colon J\to U$, and $s\in J$. From these equations,
it is evident that a strong normal frame is a normal one, the opposite
being valid
as an exception, \eg if $U$ is a neighborhood. This situation is identical
with the one for  parallel transports in $(T(M),\pi,M)$ which is a consequence
of the fact that definition~\ref{4-Defn12.2} incorporates
definition~\ref{4-Defn12.1} as its obvious special case.

	The main difference between the cases of parallel transports and
arbitrary linear transports along paths is that for the former the
condition~\eref{4-12.3} holds globally, \ie for every path $\gamma\colon
J\to M$, for some uniquely fixed $\Gamma_\mu$, while for the
latter~\eref{4-12.3} is valid, generally, locally, \ie for $\gamma\colon
J\to U$ with $U\subseteq M$, and in it $\Gamma_\mu$ are fixed but are not
uniquely defined by the transport and may depend on $U$ (see
section~\ref{4-Sect5}).
The cause for this is that
for a parallel transport, equation~\eref{4-12.3} on $M$ with uniquely
defined $\Gamma_\mu$ follows from its definition, while if for a given linear
transport  $L$ this equation holds on $U$ for some $\Gamma_\mu$, it is also
true if we replace $\Gamma_\mu$ with $\Gamma_\mu+G_\mu$ where the
matrix\ndash valued functions $G_\mu$ are subjected to the condition
$G_\mu\dot\gamma^\mu=0$ for every path $\gamma$ in $U$. If $U$ is an open
set, then $\dot\gamma(s)$ is an arbitrary vector in $T_{\gamma(s)}(M)$, which
implies $G_\mu|_U=0$, \ie in this case the 3\ndash index coefficients of $L$
are unique; just this is the case with a parallel transport when $U=M$ and
its 3\ndash index coefficients are fixed and, by definition, are equal to the
coefficients of the linear connection generating it.

	If in definition~\ref{4-Defn12.2} one replaces $D$ with a derivation
$\mathcal{D}$ along tangent vector fields and~\eref{4-12.4}
with~\eref{4-6.7}, the definition of a frame strong normal on $U$ for
$\mathcal{D}$ will be obtained. But, by proposition~\ref{4-Prop6.4}, every
frame normal on $U$ for $\mathcal{D}$ is strong normal on $U$ for
$\mathcal{D}$  and \emph{vice versa}. Therefore the concepts of a `normal
frame' and `strong normal frame', when applied to  derivations along tangent
vector fields, are identical. Returning to the considerations in
Sect.~\ref{4-Sect6}, we see that frames (strong) normal for a derivation
along tangent vector fields are \emph{strong} normal for some derivation or
linear transport along paths and \emph{vice versa}. For this reason, below
only strong normal frames for the latter objects will be investigated.

	To make the situation easier and clearer, below the following
problem will be studied. Let $(E,\pi,M)$ be a vector bundle over a $C^1$
manifold $M$, $V\subseteq M$  be an \emph{open} subset, $U\subseteq V$, and
$L$ be a linear transport along paths in $(E,\pi,M)$ whose coefficient matrix
has the form~\eref{4-12.3} on $V$, \ie for every $C^1$ path $\gamma\colon
J\to V$.%
\footnote{~%
From here follows the existence of unique 3\ndash index coefficients of $L$
on $V$ which, under a change of frames, transform into~\eref{4-5.4}. We
suppose the 3\ndash index coefficients of $L$ on $U$ to be fixed and equal to
the ones on $V$ when restricted to U.%
}
The problem to be investigated frames strong normal for $L$ on $U$.

	Let $\{e_i\}$ be a frame over $V$ in $E$ and $\{E_\mu\}$ a frame over
$V$ in $T(M)$. A frame $\{e'_i=A_i^je_j\}$ over $V$ in $E$ is strong normal
on $U\subseteq V$ if for some frame $\{E'_\mu\}$ over $V$ in $T(M)$ is
fulfilled $\Gamma'_\mu|_U=0$ with $\Gamma'_\mu$ given by~\eref{4-5.3}. Hence
$\{e'_i\}$ is strong normal on $U$ iff the matrix\ndash valued function
$A=[A_i^j]$ satisfies the \emph{(strong) normal frame equation}
	\begin{equation}	\label{4-12.6}
( \Gamma_\mu A + E_\mu(A) )|_U = 0
	\end{equation}
where $\Gamma_\mu$ are the 3-index coefficients' matrices of $L$ in
$(\{e_i\},\{E_\mu\})$.
% This equation, describing the matrix $A=[A_i^j]$ which
% provides a transition from an arbitrary to a strong normal frame(s), is also
% a consequence of~\eref{4-4.0}, \eref{4-12.3}, and the arbitrariness of
% $\gamma\colon J\to V\supseteq U$.

	If on $U$ exists a frame $\{e_i\}$ strong normal for $L$, then all
frames  $\{e'_i=A_i^je_j\}$ which are normal or strong normal on $U$ can
easily be described: for the normal frames, the matrix $A=[A_i^j]$ must be
constant on $U$ (corollary~\ref{4-Cor3.3}), $A|_U=0$, while for the strong
normal frames it must be such that $E_\mu(A)|_U=0$ for some (every) frame
$\{E_\mu\}$ over $U$ in $T(M)$ (see~\eref{4-12.6} with $\Gamma_\mu|_U=0$).

	Comparing equation~\eref{4-12.6} with analogous ones
in~\cite{bp-Frames-n+point,bp-Frames-path,bp-Frames-general}, we see that
they are identical with the only difference
that the size of the square matrices $\Gamma_1,\ldots,\Gamma_{\dim M}$, and
$A$
in~\cite{bp-Frames-n+point,bp-Frames-path,bp-Frames-general}
is $\dim M\times\dim M$ while in~\eref{4-12.6} it is
$v\times v$, where $v$  is the dimension of the vector bundle $(E,\pi,M)$,
\ie $v=\dim\pi^{-1}(x)$, $x\in M$, which is generally not equal to $\dim M$.
But this difference is completely insignificant from the view\ndash point of
solving these equations (in a matrix form) or with respect to the
integrability conditions for them. Therefore all of the results
of~\cite{bp-Frames-n+point,bp-Frames-path,bp-Frames-general},
concerning the solution of the matrix
differential equation~\eref{4-12.6}, are (\emph{mutatis mutandis}) applicable
to the investigation of the frames strong normal on a set $U\subseteq M$.

	The transferring of results
from~\cite{bp-Frames-n+point,bp-Frames-path,bp-Frames-general}
is so trivial that their explicit reformulations makes sense only if one
really needs the corresponding rigorous assertions for some concrete purpose.
For this reason, we describe below briefly the general situation and one of its
corollary.

	The only peculiarity one must have in mind, when such transferring is
carried out, consist in the observation that in this way can be obtained,
generally, only \emph{part} of the frames normal for some linear transport,
\viz the frames strong normal for it. But such a state of affairs is not a
trouble as we need a single normal frame to construct all of them by means of
corollary~\ref{4-Cor3.3}.

	If $\gamma_n\colon J^n\to M$, $J^n$ a neighborhood in
$\mathbb{R}^n$, $n\in\mathbb{N}$, is a $C^1$ injective map,
then~\cite[theorem~3.1]{bp-Frames-general} a necessary and sufficient
condition for the existence of frame(s) strong normal on $\gamma_n(J^n)$
for some linear transport along paths or derivation along paths or along
vector fields tangent to $M$, is in some neighborhood (in $\mathbb{R}^n$) of
every $s\in J^n$ their (3\ndash index) coefficients to satisfy the equations
	\begin{equation}	\label{4-12.7}
\bigl(
  R_{\mu\nu} (-\Gamma_1\circ\gamma_n,\ldots,-\Gamma_{\dim M}\circ\gamma_n)
\bigr)
	(s)
  = 0,
\qquad \mu,\nu=1,\ldots,n
	\end{equation}
where $R_{\mu\nu}$ are given via~\eref{4-5.6} for $x^\mu=s^\mu$,
 $\mu,\nu=1,\dots,n$ with $\{s^\mu\}$ being Cartesian coordinates in
$\mathbb{R}^n$.

	From~\eref{4-12.7} an immediate observation
follows~\cite[sect.~6]{bp-Frames-general}: strong normal frames always exist
at every point ($n=0$) or/and along every $C^1$ injective path ($n=1$).
Besides, these are
the \emph{only cases} when normal frames \emph{always exist} because for them
~\eref{4-12.7} is identically valid. On submanifolds with dimension greater
than or equal to two normal frames exist only as an exception if (and only
if)~\eref{4-12.7} holds. For $n=\dim M$ equations~\eref{4-12.7} express the
flatness of the corresponding linear transport~\cite{bp-LTP-Cur+Tor} or
derivation~\cite[sect.~2]{bp-Frames-path} to which we shall return to elsewhere.
% (see Sect.~\ref{4-Sect7}, equation~\eref{4-7.13}, or
% corollary~\ref{4-Cor7.1}) or, if $n=\dim M$, derivations (see
% Sect~\Ref[3]{Sect2}).

	It is almost evident, in the coordinates used,
equation~\eref{4-12.7} is identical with~\eref{4-5.14} for
$N=\gamma_n(J^n)$ and $p=\gamma_n(s)$. Thus, on a submanifold or along
injective mappings, the existence of normal frames (for linear transports of
the considered type) implies the existence of strong normal frames.

\section {Conclusion}
\label{4-Conclusion}

	In the preceding sections we have developed the generic theory of
linear transports along paths in vector bundles and of frames normal
for them and for derivations along paths and/or along tangent vector fields
(if the bundle's base is a manifold in the last case). Below we make some
conclusions from the material presented and point out links with other
results in this field.

	From proposition~\ref{4-Prop5.1} and theorem~\ref{4-Thm5.2}, we
know that only linear transports/de\-ri\-va\-tions along paths with (2-index)
coefficients given by~\eref{4-5.1} admit normal frames. Besides, from
equations~\eref{4-5.1} and~\eref{4-5.3}, it follows that frames normal on a
subset $U\subseteq M$ for such transports/derivations along paths exist if
and only if the matrix differential equation
	\begin{equation}	\label{4-11.21}
\left.\left[ \dot\gamma^\mu\left(
\Gamma_\mu A + \frac{\pd A}{\pd x^\mu} 	\right) \right]\right|_U
   = 0
\end{equation}
has a solution for every $\gamma\colon J\to U$ with respect to $A$.%
\footnote{%
If such $A$ exist in a frame $\{e_i\}$, then the frame
$\{e_i^{\prime}=A_{i}^{j}e_j\}$ is normal on $U$ and vice versa;
see~\eref{4-5.3} and proposition~\ref{4-5.2}.%
}
In fact, the equations~\eref{4-5.14} are the integrability conditions
for~\eref{4-11.21}.%
\footnote{%
If~\eref{4-5.5} hold and $U$ is a neighborhood, then
$A=Y(p,p_0;-\Gamma_1,\ldots,-\Gamma_{\dim M}) A_0$, $A_0$ being
non\ndash degenerate matrix. %
}
Evidently, the same is the situation with
derivations along tangent vector fields (see Sect.~\ref{4-Sect6}) when, due
to~\eref{4-6.5}, such a derivation admits frames normal on $U$ iff the
equation
	\begin{equation}	\label{4-11.22}
\bigl(\Gamma_X A + X(A) \bigr)\big|_U = 0,
	\end{equation}
${\Gamma}_X$ being the derivation's matrix along a vector field $X$,
has a solution with respect to $A$. As we proved in Sect.~\ref{4-Sect6},
if $X$ is arbitrary and tangent to the paths in $U$, this equation is
equivalent to~\eref{4-11.21} with $\Gamma_\mu$ being the matrices of the
3\ndash index coefficients of the derivation; if $X$ is completely
arbitrary,~\eref{4-11.22} is equivalent to equation~\eref{4-11.23} below.

	Now it is time to recall that, from a mathematical view-point, the
series of papers~\cite{bp-Frames-n+point,bp-Frames-path,bp-Frames-general} is
actually devoted precisely to the solution of the equation%
\footnote{%
In~\cite{bp-Frames-n+point,bp-Frames-path,bp-Frames-general}
the notation $W_X$ instead of $\Gamma_X$ is used.%
}
	\begin{equation}	\label{4-11.23}
\left.\left(
\Gamma_\mu A + \frac{\pd A}{\pd x^\mu} 	\right)\right|_U
   = 0
\end{equation}
which is equivalent to~\eref{4-11.21} if $U$ is a neighborhood.
The general case is explored in~\cite{bp-Frames-general},
while~\cite{bp-Frames-path} investigates the case $U=\gamma(J)$ for
$\gamma\colon J\to M$ and~\cite{bp-Frames-n+point} is concentrated on the
one in which $U$ is a single point or a neighborhood in $M$.
The fact that in the
works mentioned are studied frames normal for derivations of the tensor
algebra over a manifold $M$ is inessential because the \emph{equations
describing the matrices} by means of which is performed the transformation
from an arbitrary frame to a (strong) normal one are the same in these
papers and in the present investigation. The only difference is \emph{what
objects} are transformed by means of the matrices satisfying~\eref{4-11.22}:
in the present work these are the frames in the restricted bundle space
$\pi^{-1}(U)\subseteq E$, while in the above series of works they are the
tensor bases over $U$, in particular the ones in the bundle tangent to $M$.
In~\cite{bp-Frames-n+point,bp-Frames-path,bp-Frames-general}
the only explicit use of the derivations of the tensor algebra over $M$ was
to define their components (2\nobreakdash-index coefficients) and the
transformation law for the latter. Since this
law~\cite[equation~(2.2)]{bp-Frames-general} is identical with~\eref{4-6.5},%
\footnote{%
The transformation laws~\eref{4-2.26} and~\eref{4-5.3} can be considered, under
certain conditions, as special cases of~\eref{4-6.5}.%
}
all  results concerning the 2- and 3\nobreakdash-index coefficients of
derivations of the tensor algebra over $M$ and the ones of derivations
along tangent vectors in vector bundle $(E,\pi,M)$ \emph{coincide}.

	Thus, we have came to a very important conclusion:
\emph{all of the results
of~\cite{bp-Frames-n+point,bp-Frames-path,bp-Frames-general}
concerning S\ndash derivations, their components, and frames normal for them
are mutatis mutandis valid (as investigated in the present work)
for  linear transports along
paths, derivations along paths or along tangent vector fields, their
coefficients (or components), and the frames (strong) normal for them in
vector bundles with a differentiable manifold as a base}.
The only change, if required, to transfer the results is to replace the term `S\ndash
derivation' with `derivation along tangent vector fields', or `derivation
along paths', or `linear transport along paths'
and, possibly, the term `normal frame' with `strong normal frame'.

	Because of the widespread usage of covariant derivatives (linear
connections), we want to mention them separately regardless of the fact that
this case was completely covered
in~\cite{bp-Frames-n+point,bp-Frames-path,bp-Frames-general}. As a
consequence of~\eref{4-2.30}, the covariant derivatives are linear
derivations on the whole base $M$ (as well as on any its subset). Thus for
them the condition~\eref{4-5.1} is identically satisfied. Therefore, by
theorem~\ref{4-Thm5.2}, a covariant derivative (or the corresponding
parallel transport) admits normal frames on a submanifold $U\subseteq M$
iff~\eref{4-5.14} holds on $U$.
Consequently, every covariant derivative admits normal frames at every
point or along any given smooth injective path. However, only the flat covariant
derivatives on $U$ admit frames normal on $U$ if $U$ is a neighborhood ($\dim
U=\dim M$).  Further general details concerning this important case can be
found in~\cite[sect.~5]{bp-Frames-general}.

	In theoretical physics, we  find applications of a number of linear
transports along paths~\cite{bp-LT-Deriv-tensors}:
parallel~\cite{K&N-1,Schouten/Ricci},
Fermi-Walker~\cite{Hawking&Ellis,Synge},
Fermi~\cite{Synge},
Truesdell~\cite{Walwadkar,Walwadkar&Virkar},
Jaumann~\cite{Radhakrishna&et_al.},
Lie~\cite{Hawking&Ellis,Schouten/Ricci},
modified Fermi\ndash Walker and
Frenet\ndash Serret~\cite{Dandoloff&Zakrzewski}, etc.
	Our results are fully applicable to all of
them (see~\cite[proposition~4.1]{bp-LT-Deriv-tensors}), in particular for all
of them there exist frames normal at a given point or/and along smooth
injective paths.

	We end with a few words about gravity. A comprehensive analysis,
based on~\cite{bp-Frames-n+point,bp-Frames-path,bp-Frames-general}, of the
connections between gravity and normal frames is given in~\cite{bp-PE-P?}. The
importance of the concept of `normal frame' for physics comes from the
fact that it is the mathematical object representing the physical concept of
an `inertial frame'. Moreover, in~\cite{bp-PE-P?} we proved that the (strong)
equivalence principle is a theorem according to which these two types of
frames coincide.  Thus, we hope, the present investigation may find
applications in the further exploration of gravity.

	The formalism developed in the present work can find natural
application in gauge theories~\cite{bp-NF-D+EP},dfields, which mathematically
are linear connections, to whose coefficients our results are applicable.
% An example of
% such an application is presented in the Appendix below.

\section*{Acknowledgments}

% 	The referee of this work has made a great number of useful
% suggestions concerning the way of presenting  the material, the English
% grammar of the text as well as making some statements more clear, precise or
% rigorous. The author thanks him/her for the efforts and time spent in critical
% reading of the manuscript and helping toward the improvement of its final form.
	The author expresses his gratitude to Professor James D.~Stasheff
(Math-UNC, Chapel Hill, NC, USA) for his interest in the present paper and
help for correcting the English of the manuscript.

% $$$$$$$$$$$$$$$$$$$$$$$$$$$$$$$$$$$$$$$$$$$$$$$$$$$$$$$$$$$$$$$$$$$$$$$$$
% $$$$$$$$$$$$$$$$$$$$$$$$$$$$$$$$$$$$$$$$$$$$$$$$$$$$$$$$$$$$$$$$$$$$$$$$$

%	BIBLIOGRAPHY
\addcontentsline{toc}{section}{References}
\bibliography{bozhopub,bozhoref}
\bibliographystyle{unsrt}
 \addcontentsline{toc}{subsubsection}{This article ends at page}

% FINAL design of the INDEX

%	BEGINNING of the "see" index entries  ###############=====>>>
% \index{|see{}}
 \index{transport|see{transport along paths or maps,
	linear transport along paths}}

 \index{differentiation|see{derivation}}
 \index{derivation|see{derivation along paths
	or tangent vector fields,
	linear derivation along tangent vector fields}}

 \index{basis|see{frame}}
 \index{frame!normal|see{normal frame}}
 \index{coordinates!normal|see{normal frame}}
%	END of the "see" index entries <<<=====###############

%	SUBJECT INDEX
%  \addcontentsline{toc}{section}{Index}
% \printindex

\end{document}